\documentclass[fleqn,usenatbib]{mnras}

\usepackage{newtxtext,newtxmath}
\usepackage[normalem]{ulem}
\usepackage[T1]{fontenc}

\usepackage{longtable}

\DeclareRobustCommand{\VAN}[3]{#2}
\let\VANthebibliography\thebibliography
\def\thebibliography{\DeclareRobustCommand{\VAN}[3]{##3}\VANthebibliography}

\usepackage{xspace}
\usepackage{multirow}
\usepackage{makecell}
\usepackage{array,calc}

\newcommand{\iprocess}{$i$-process\xspace}
\newcommand{\sprocess}{$s$-process\xspace}
\newcommand{\rprocess}{$r$-process\xspace}
\newcommand{\cisotope}{$\rm^{12}C/^{13}C$\xspace}
\newcommand{\mgtwentyfour}{$\rm^{24}Mg$\xspace}
\newcommand{\mgtwentyfive}{$\rm^{25}Mg$\xspace}
\newcommand{\mgtwentysix}{$\rm^{26}Mg$\xspace}

\usepackage{graphicx}	
\usepackage{amsmath}	

\title[Searching for the $i$-process in Barium Stars]{Searching for Signatures of $i$-process Nucleosynthesis in Metal-Rich, Lead-rich Barium Stars\thanks{Based on observations taken with Gemini/GHOST (GS-2026A-FT-104), McDonald/Tull, and European Southern Observatory/UVES (113.26NB.001).}}

\author[C. Manea et al.]{
Catherine Manea,$^{1}$\thanks{E-mail: catherinemanea@gmail.com}\thanks{NSF Astronomy and Astrophysics Postdoctoral Fellow}
Madeleine McKenzie,$^{2}$\thanks{NASA Hubble Fellow}
Gail Zasowski,$^{1}$
Giulia C. Cinquegrana,$^{3,4}$
Amaya Sinha,$^{5}$
\newauthor
Sarah G. Kane,$^{6}$
Jaden Levine,$^{7}$
Ellie Mebius$^{1}$
and Aleksandra Winkler$^{1}$
\\
$^{1}$Department of Physics \& Astronomy, University of Utah, Salt Lake City, UT 84112, USA\\
$^{2}$The Observatories of the Carnegie Institution for Science, 813 Santa Barbara Street, Pasadena, CA 91101, USA\\
$^{3}$Centre for Astrophysics \& Supercomputing, Swinburne University of Technology, Hawthorn VIC 3122, Australia\\
$^{4}$School of Physics \& Astronomy, Monash University, Clayton VIC 3800, Australia\\
$^{5}$Space Telescope Science Institute, Baltimore, MD 21218, USA\\
$^{6}$Institute of Astronomy, University of Cambridge, Madingley Road, Cambridge CB3 0HA, UK\\
$^{7}$Department of Astronomy, New Mexico State University, Las Cruces, NM 88003, USA
}

\date{Accepted XXX. Received YYY; in original form ZZZ}

\pubyear{\the\year{}}

\begin{document}
\label{firstpage}
\pagerange{\pageref{firstpage}--\pageref{lastpage}}
\maketitle

\begin{abstract}
The intermediate neutron-capture process (\iprocess) has been invoked to explain the peculiar compositions of carbon-enhanced metal-poor (CEMP) stars enriched in both slow- ($s$) and rapid- ($r$) neutron-capture elements. Recent literature has questioned its operation at higher metallicities, but observational evidence remains limited.  We analyse Gemini/GHOST and McDonald/Tull spectra of six $\rm -0.7 < [Fe/H] < -0.1$ stars identified in GALAH by their extreme Ba and Eu (or Sm) compositions ($\gtrsim$10$\times$ solar). We also reanalyse archival spectra of two literature \iprocess candidates. We determine abundances in up to 42 elements (Li through Pb), \cisotope, and Mg isotopic ratios, and compare to literature barium and CEMP stars and multiple asymptotic giant branch (AGB) model grids.  Three stars exhibit extreme [Pb/Fe] (+1.79$\pm$0.19 to +2.50$\pm$0.32~dex), and two extreme [Eu/Fe] (+1.10$\pm$0.11 to +1.30$\pm$0.11~dex), distinguishing them from typical metal-rich barium stars.  Two of these three exhibit abundance patterns that are inconsistent with \sprocess models and best explained by existing \iprocess predictions.  The remaining sample is either consistent with \sprocess enrichment or suggestive of atomic diffusion. Both literature candidates are well-fit by the $i$-process model, but conventional \sprocess enrichment cannot be excluded. Our findings reinforce that metal-rich AGB stars contribute to the production of elements classically attributed to the \rprocess. These results identify promising candidates for \iprocess enrichment near solar metallicity and motivate additional \iprocess models to understand the contribution of this nucleosynthetic process to the chemical evolution of the Milky Way disk.
\end{abstract}




\section{Introduction} 
\label{sec:intro}

Constraining the astrophysical origins of the elements is essential for understanding the conditions that govern nucleosynthesis and for developing realistic models of stellar and Galactic chemical evolution \citep{Burbidge1957, Busso1999, Kobayashi2020}.  Despite substantial progress in identifying the astrophysical sites and processes responsible for nucleosynthesis, the origins of many of the heavy elements remain uncertain. Elements heavier than Fe form primarily through neutron capture onto heavy seed nuclei (typically $^{56}$Fe), where the resulting elemental distribution is then governed by factors such as the density of the neutron stream \citep[e.g.,][]{Kappeler2011}. 

Historically, research has focused on the two extremes of the neutron-density spectrum. High neutron densities ($>\rm 10^{20},cm^{-3}$) facilitate the rapid neutron-capture ($r$-) process, during which successive neutron captures can occur on timescales shorter than those of \(\beta\)-decay. This drives nucleosynthesis away from the valley of stability \citep[e.g.,][]{Thielemann2026}. The $r$-process has been confirmed to occur in neutron-star mergers \citep[e.g.,][]{Cote2018}, with other sites proposed to occur in extremely energetic supernovae (e.g., magnetorotational supernovae \citealp[e.g.,][]{Winteler}, collapsars \citealp[e.g.,][]{Siegel2019}).  Lower neutron densities ($\rm 10^{7} - 10^{11}~cm^{-3}$) lead to the slow neutron-capture process (\sprocess), restricting production to nuclides along the valley of stability \citep[e.g.,][]{Gallino1998, Cristallo2009, Karakas2014, Lugaro2023ARNPS..73..315L, Lugaro2025}. The strong s-process channel occurs in thermally pulsing asymptotic giant branch (AGB) stars \citep[e.g.,][]{Herwig2005, Karakas2014, Lugaro2025} via a radiative $^{13}$C pocket that releases neutrons via $^{13}$C($\alpha$,n)$^{16}$O during interpulse periods and a $^{22}$Ne($\alpha$,n)$^{25}$Mg source during thermal pulses (for a review, see \citealt{Lugaro2023ARNPS..73..315L}). Weak s-process contributions are also made in massive stars\citep[e.g.,][]{Pignatari2010}. Between these two well-studied neutron density regimes lies the intermediate neutron-capture process (\iprocess), which occurs at neutron densities of $\sim \rm 10^{14} - 10^{16} cm^{-3}$ \citep[][]{Cowan1977ApJ...212..149C}. However, despite growing evidence for its contribution to heavy-element nucleosynthesis (most famously, Sakurai's object, \citealp{Nakano1996}), the astrophysical origins of the \iprocess remain an outstanding mystery.
 
One of the leading proposed sites for the $i$-process are proton ingestion events (PIEs) \citep[e.g.,][]{Campbell2008, Cristallo2009, Campbell2010, Herwig2011, Lugaro2023ARNPS..73..315L, Wiedeking2025}. PIEs are thought to occur during the core helium flash in low-mass stars and during the first few thermal pulses of the AGB phase in intermediate-mass stars. During these episodes, hydrogen-rich material is engulfed into convective helium-burning regions, leading to the production of $^{13}$C and the subsequent release of neutrons via $^{13}$C($\alpha$,n)$^{16}$O. These neutron streams can produce substantially higher densities than the radiative \(^{13}\)C pocket, and sustain these high densities for longer than the brief bursts associated with the \(^{22}\)Ne source during thermal pulses.
 
The occurrence of PIEs in AGBs has traditionally been associated with very low-metallicity regimes. In these environments, the scarcity of CNO nuclei limits hydrogen burning through the CNO cycle, favouring the less temperature-sensitive pp-chains. This can result in a comparatively weak hydrogen-burning shell and a reduced entropy barrier between the hydrogen-rich envelope and the underlying helium-burning region. It has therefore been proposed that convection driven by a helium flash can extend into the overlying hydrogen-rich layers, triggering a PIE. This metallicity association is also reflected in the observational literature, where searches for \iprocess-enriched stars have largely focused on the metal-poor regime. Accordingly, theoretical investigations of PIE-driven \iprocess nucleosynthesis have predominantly considered extremely metal-poor stars. However, the occurrence of PIEs is sensitive to the treatment of mixing at convective boundaries, which is one of the most uncertain features of stellar evolution modelling. Different prescriptions for convective boundary mixing can thus lead to substantially different predictions for the conditions under which proton ingestion occurs, and potentially allow PIEs to operate at higher metallicities.

For example, \citet[][hereafter K23]{Karinkuzhi2023} present two $2\,\mathrm{M}_{\odot}$ AGB models at $\rm [Fe/H] = -0.5$, adopting standard mixing assumptions in one model and additional mixing in the other. The latter produces a PIE, reaching neutron densities characteristic of the \iprocess. K23 compare their models with three stars at approximately $\rm [Fe/H] \sim -0.5$, which exhibit enhancements in both barium (associated with the $s$-process) and europium (associated with the $r$-process), at levels approximately ten times solar. They find that the observed abundance patterns are better reproduced by their \iprocess model than by a conventional $s$-process model. These results suggest that \iprocess nucleosynthesis may not be restricted to the metal-poor regime, motivating further investigation across a broader range of metallicities. It therefore remains unclear whether the apparent prevalence of \iprocess signatures at low metallicity reflects an intrinsic preference for these environments or, at least in part, the metallicity distribution of the stars studied to date. Extending observational searches to more metal-rich stars will help establish the metallicity dependence of \iprocess nucleosynthesis and constrain the stellar conditions under which it can occur. 

Barium stars are low-mass binary companions enriched by mass transfer from a former AGB star, now a white dwarf, making them valuable probes of AGB nucleosynthesis and its associated \(s\)- and \(i\)-processes. \citep[e.g.,][]{McClure1980, Smiljanic2007, Pereira2011, deCastro2016, Karinkuzhi2018, Roriz2021, Cseh2022, Vilagos2024}. The recent discovery of nearly 3,000 barium stars by \citet{Levine_2026_BaStars} in the third data release of the GALAH survey \citep{GALAHDR3} has enabled us to search for \iprocess signatures among a larger sample.  In this work, we follow up six metal-rich ($\rm [Fe/H] > -1$) barium stars with GALAH-reported [Ba/Fe] and [Eu/Fe] exceeding 10$\times$ solar to assess if \iprocess nucleosynthesis can explain their extreme abundance patterns.  We also reanalyse spectra of the K23 barium stars to re-test their classifications.  We conduct the most extensive chemical analysis of metal-rich, candidate \iprocess-enriched barium stars yet, determining  \cisotope, \mgtwentyfour, \mgtwentyfive, and \mgtwentysix ratios, and elemental abundances in up to 42 elements from Li through Pb, and compare to the literature and existing $i$- and $s$-process models where available. 
We identify two new stars whose abundance patterns, characterised most notably by extreme Eu and Pb enhancements, cannot be reproduced by any of the tested \sprocess models but are well matched by existing \iprocess predictions. These findings provide new evidence for \iprocess nucleosynthesis at near-solar metallicities and motivate further investigation of its role in metal-rich stellar populations.

\section{Data}
\label{sec:data}


\begin{table*}
\centering
\setlength{\tabcolsep}{4.5pt}
\caption{Description of sample.}
\label{tab:observations}
\begin{tabular}{c|lcc|ccccc|>{\centering\arraybackslash}m{\widthof{2024-07-04}}cc}
\hline
 & & & & \multicolumn{5}{c|}{GALAH-reported} & \multicolumn{3}{c}{Observations} \\
 & Source & $V$ & RUWE & $T_{\rm eff}$ & $\log g$ & [Fe/H] & [Ba/Fe] & [Eu/Fe] & UT Date & $t_{\rm exp}$ & SNR pix$^{-1}$ \\
 & & (mag) & & (K) & (dex) & (dex) & (dex) & (dex) & & (s) & (4820\,\AA) \\
\hline
\hline
\multirow{8}{*}[-4.5ex]{\rotatebox[origin=c]{90}{\small\makecell{New \iprocess\\candidates}}} &
  \multicolumn{11}{l}{\textit{Gemini-South/GHOST (program GS-2026A-FT-104)}} \\
 & UCAC4 278-068826 & 13.592 & 3.081 & 4849 & 2.98 & $-0.65$ & 2.12 & 1.06 & 2026-02-26 & 2130 & 41 \\
 & UCAC4 179-210559 & 13.970 & 2.413 & 5220 & 3.92 & $-0.46$ & 1.06 & 1.02 & 2026-02-26 & 1900 & 34 \\
 & UCAC4 265-014129 & 13.170 & 1.388 & 6178 & 3.49 & $-0.22$ & 1.16 & 1.27 & 2026-02-27 & 690 & 27 \\
 & UCAC4 445-009460 & 14.145$^{a}$ & 2.724 & 5557 & 4.70 & $-0.69$ & 1.80 & 1.35 & 2026-03-18 & 2200 & 42 \\
\cline{2-12}
 & \multicolumn{11}{l}{\textit{McDonald 2.7\,m/Tull coud\'e spectrograph}} \\
 & HD 216387      & 10.35 & 4.767 & 6305 & 4.14 & $-0.19$ & 1.10 & --$^{b}$ & 2026-10-12 & 5400 & 44 \\
 & TYC 1238-203-1 & 11.14 & 1.764 & 5061 & 3.34 & $-0.12$ & 1.06 & --$^{c}$ & 2026-10-12 & 14400 & 19 \\
\hline
\hline
\multirow{6}{*}[-2.5ex]{\rotatebox[origin=c]{90}{\small\makecell{Literature\\candidates}}} &
  & & & \multicolumn{5}{c|}{Literature-reported$^{d}$} & \multicolumn{3}{c}{Observations$^{e}$} \\
\cline{2-12}
 & \multicolumn{11}{l}{\textit{VLT/UVES (ESO Science Archive, program 113.26NB.001)}} \\
 & TYC 2250-1047-1 & 11.67 & 1.294 & 5335 & 3.71 & $-0.55$ & 1.17 & 0.88 & 2024-07-04\newline 2024-07-08 & 5400 & 38 \\
 & TYC 591-1090-1  & 12.08 & 1.387 & 5267 & 3.68 & $-0.30$ & 1.12 & 0.93 & 2024-06-12\newline 2024-07-03 & 7100 & 24 \\
\hline
\end{tabular}
\begin{flushleft}
\footnotesize
\textit{Notes.} RUWE values are from \textit{Gaia} DR3.\\
$^{a}$We could not find measurements of $V$, so we report $R$.\\
$^{b}$Selected due to high GALAH-reported [Sm/Fe] (0.88\,dex).\\
$^{c}$Selected due to high GALAH-reported [Sm/Fe] (0.92\,dex).\\
$^{d}$$T_{\rm eff}$, $\log g$, [Fe/H] and [Ba/Fe] for candidate \iprocess\ stars from \citet{Karinkuzhi2021b}; [Eu/Fe] from \citet{Karinkuzhi2023}.\\
$^{e}$Exposure times are summed over both visits.
\end{flushleft}
\end{table*}

\subsection{Sample Selection and Spectra}\label{sec:sample}

Our sample comprises eight candidate \iprocess-enriched stars, including six newly identified stars and two previously reported literature candidates.  Spectra of these targets were collected from three different facilities.  We describe our target selection below and present our final sample in Table \ref{tab:observations}.  Details of our observational setup and data reduction are included in Appendix \ref{appendix:obs}.

\begin{figure}
    \centering
    \includegraphics[width=.9\linewidth]{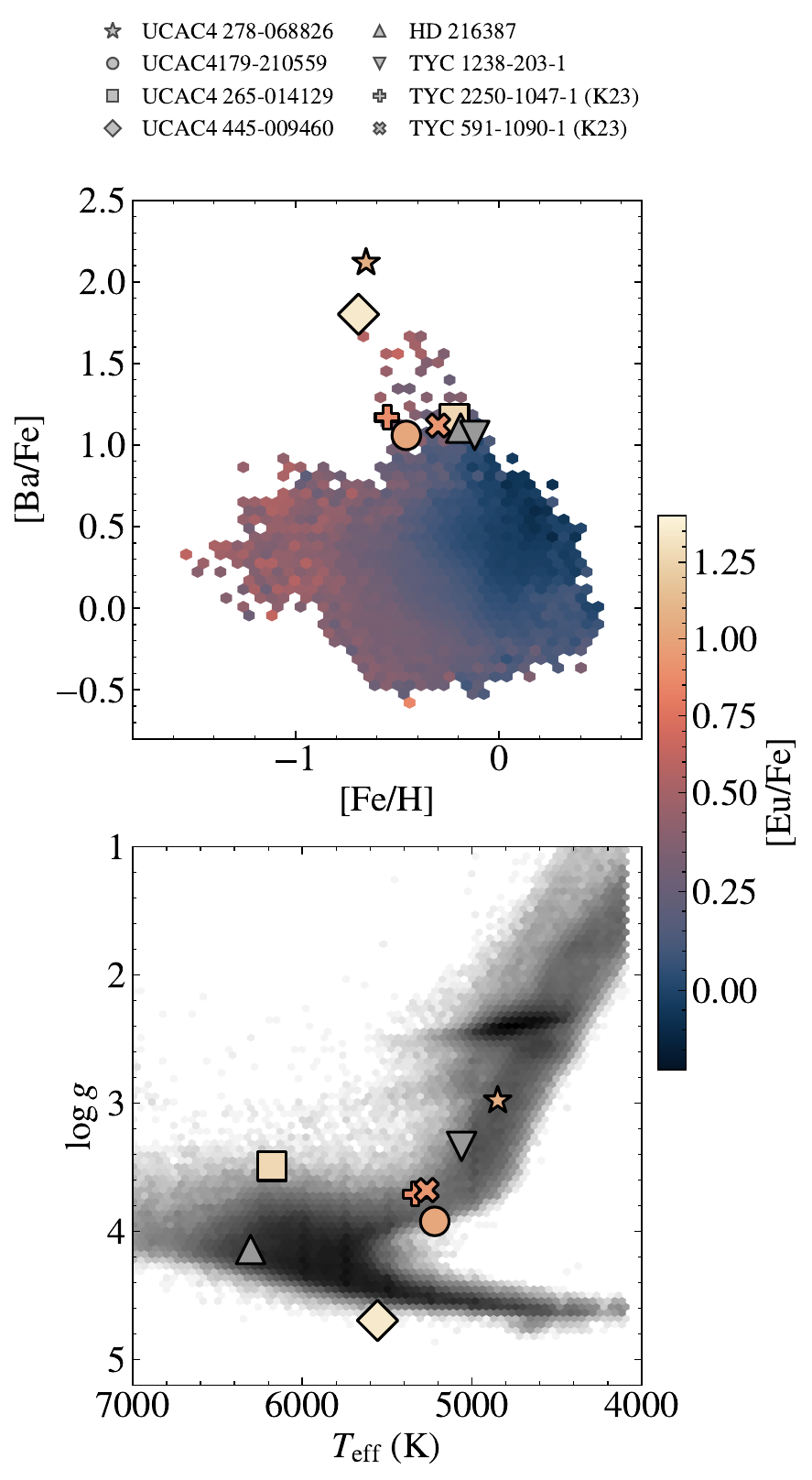}
    \caption{\textit{Top:} Literature-reported [Ba/Fe] versus [Fe/H] for the candidate \iprocess-enriched stars followed up in this study (markers), colored by their reported [Eu/Fe]. Gray markers have no reported [Eu/Fe] but exhibit enhanced [Sm/Fe] (Section~\ref{sec:sample}). The background shows GALAH DR3 field stars, binned and colored by median [Eu/Fe] using the same scale. \textit{Bottom:} Kiel diagram of the same stars over the GALAH DR3 field.}
    \label{fig:sample}
\end{figure}

We searched for candidate metal-rich $i$-process-enriched stars in GALAH, starting from the high-fidelity (i.e., unflagged, 4500~K~\textless~T$_{\rm eff}$~\textless~6500~K) barium star sample of \citet{Levine_2026_BaStars} and selecting all stars with GALAH DR3-reported [Ba/Fe] and [Eu/Fe]  \textgreater+1 dex.  We did not impose any criteria on their relative [Eu/Ba] ratios.  This selection returned 36 stars.  The additional criterion of $\rm [Fe/H] > -1$ reduced the sample to 20~stars.

We sought Gemini observations through the Fast Turnaround process, which accepts proposals on a monthly basis and allocates limited amounts of time depending on scheduling availability.  To maximise chances of success, we limited our sample to just four stars, intentionally selecting targets that span the Kiel diagram (Figure \ref{fig:sample}).  Because we were interested in candidate \iprocess stars enriched by AGB companions, we required all to have Gaia RUWE $\geq$ 1.4, suggestive of binarity \citep[see discussion in, e.g.,][]{Levine_2026_BaStars}.  Spectra were then collected with Gemini South's GHOST spectrograph \citep{Kalari2024}.

In addition to Gemini observations, we collected spectra of two more candidates during an October 2026 run at McDonald Observatory.  No stars with GALAH-reported [Ba/Fe] $> 1$ and [Eu/Fe] $> 1$ were observable during this run. Instead, we selected two barium stars with high GALAH [Sm/Fe] ($\sim 0.9$). Like Eu, Sm has significant contributions from the \rprocess in the Solar System (65\% $r$; \citealt{Prantzos2020}), so a strong Sm enhancement may be suggestive of \iprocess enrichment. Spectra were collected with the Tull spectrograph on the Harlan J. Smith 2.7m telescope \citep{Tull1995}.

Beyond investigating new candidate \iprocess-enriched stars identified in GALAH, we seek to validate our methods and reassess the classifications of the existing solar-metallicity candidate \iprocess-enriched stars from K23.  K23 determined abundances for their sample from HERMES/Mercator spectra of their targets. At the time of this writing, the HERMES archive was not functioning, so we could not obtain their spectra.  However, they also observed two of their three stars with UVES \citep{UVES}, which we analysed as part of this work.

Using orbital energies and angular momenta from GALAH DR3's dynamics value-added catalogue \citep{GALAHDR3}, we verified that all eight stars exhibit dynamics consistent with disc membership.  This makes it unlikely that their contributions reflect the unique $r$-process chemical evolution of accreted systems \citep[e.g.,][]{Monty2024}.  \citet{Kane2026b} notably identified a sample of 17 vetted Eu-enhanced disk stars in GALAH, a subset of them Ba-enhanced.  None of our stars fall in their sample, though that sample is also prime for searching for \iprocess signatures.

\subsection{Fundamental Parameters, Elemental Abundances, and Isotopic Ratios}
\begin{figure*}
    \centering
    \includegraphics[width=.9\linewidth]{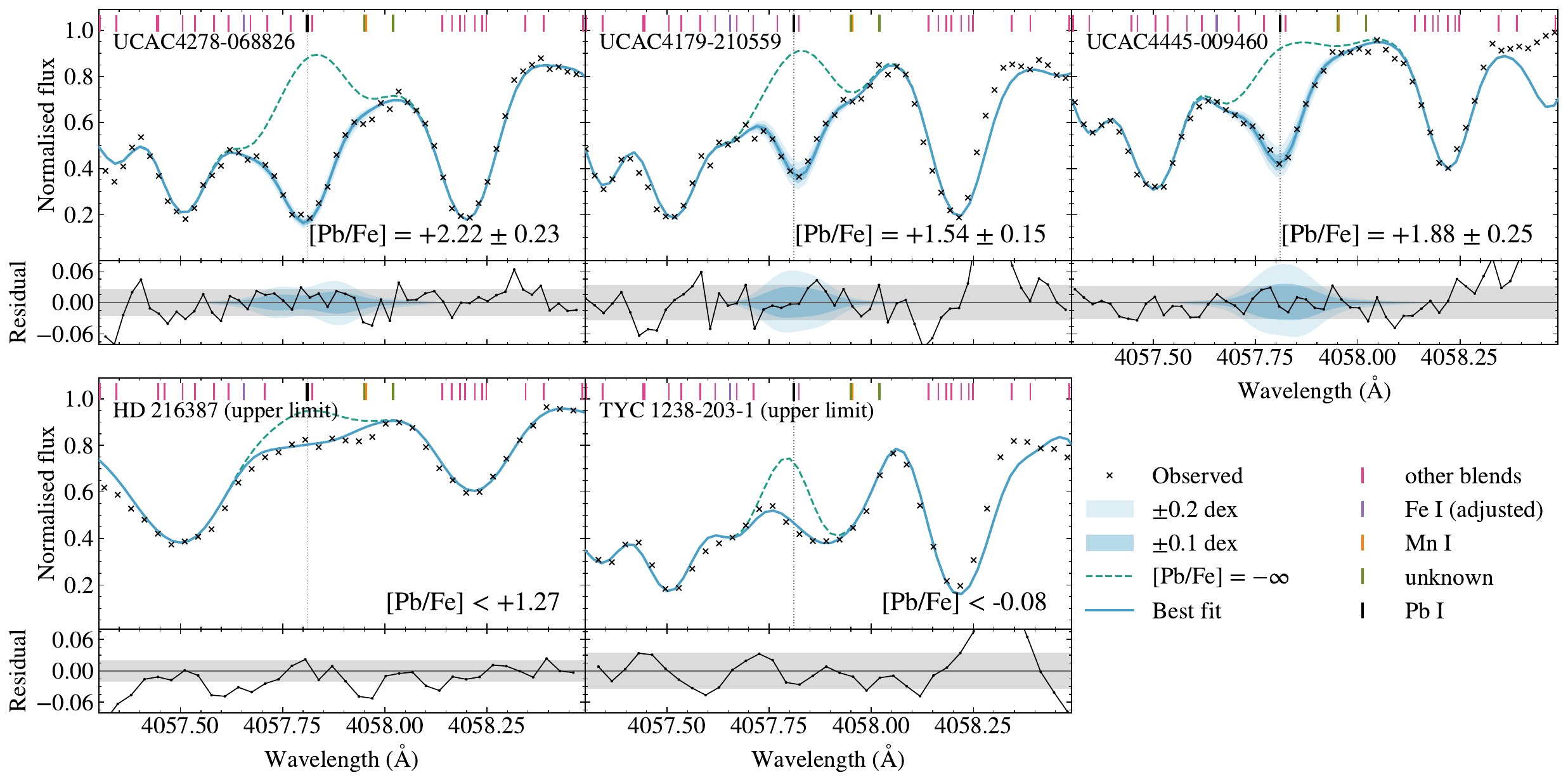}
    \caption{Synthetic spectral fits (blue solid line) to the observed Pb \textsc{i} $\lambda$4057.8 \AA~feature (black crosses) in the five stars in our sample for which we can report abundances (first row) or upper limits (second row). Shading illustrates the spectral sensitivity to abundance variations of $\pm$0.1 and $\pm$0.2 dex. Here, the printed abundances reflect the LTE best-fit value. The extreme [Pb/Fe] compositions of this sample make it possible to estimate abundances using the $\lambda$4057.8 \AA~feature despite blending with surrounding features that is often prohibitive.}
    \label{fig:pb}
\end{figure*}

\subsubsection{Parameters and Elemental abundances}

To maximize internal consistency and ensure spectroscopic equilibrium, we re-derive the fundamental parameters of the GHOST and Tull stars before measuring line-by-line abundances of 42 elements spanning the light (Li, C, N, O), odd-$Z$ (Na, Al, K), $\alpha$ (Mg, Si, S, Ca, Ti), iron-peak (Sc, V, Cr, Mn, Fe, Co, Ni, Cu, Zn\footnote{Cu and Zn are also considered neutron-capture elements given their production via the weak $s$-process in massive stars \citep[e.g.,][]{Pignatari2010}}) and neutron-capture elements, from the first peak (Rb, Sr, Y, Zr, Nb, Mo, Rh) through the second peak and lanthanides (Ba, La, Ce, Pr, Nd, Sm, Eu, Gd, Tb, Dy, Er, Lu) to the third peak (Hf, W, Os, Pb).  For the K23 stars, we adopt their reported fundamental parameters before re-deriving abundances.  We describe our procedure for determining parameters and abundances with \texttt{BACCHUS}, a Turbospectrum-based spectral synthesis fitting code, and associated uncertainties in Appendix \ref{appendix:abunds}. Our line selection is listed in Table \ref{tab:lineselection}.

Pb is a particularly important element for our analysis as it is predicted to be a discriminant between $i$- and $s$-process nucleosynthesis \citep[e.g.,][]{VanEck2003, Lugaro2015, Hampel2019, Karinkuzhi2023}.  As such, we perform a careful by-hand analysis of the Pb \textsc{i} $\lambda$4057.8~\AA~feature with Turbospectrum \citep{Plez2012}, fixing all parameters and abundances to those determined with \texttt{BACCHUS} and identifying the Pb abundance that minimises the $\chi^2$ between the data and synthesis.  We use \texttt{linemake} \citep{linemake} to create a line list for this region, including molecular lines (CH, C$_2$, CN, TiO, SiH) and hyperfine structure (HFS)
and isotope splitting for Pb \textsc{i} \citep{Roederer2012}, V \textsc{i}, Mn \textsc{i}, and Co \textsc{i}. We then checked the list against laboratory data. The \ion{Mg}{i} 4057.505~\AA\ line was set to $\log gf = -0.90$ following NIST.  Furthermore, the \ion{Fe}{i} 4057.655~\AA\ line, which has no laboratory measurement in NIST and is a calculated value, was overestimated in all sample spectra and in the \citet{Hinkle2000} Arcturus atlas.  We adopted $\log gf = -2.14$ for this line, which better fit the spectra of our sample and Arcturus and better anchored the blue wing of the Pb feature.  Furthermore, an unknown feature near the neighbouring red Mn \textsc{i} $\lambda$4057.9\AA~feature appears in all our spectra.  This feature doesn't contribute significant flux to the Pb line, but it does anchor its red wing. We add two additional lines of a generic unknown element, and during fitting, we varied the abundance of this unknown element to better fit the redward region and avoid overestimating the Pb abundance. We present our final Pb region line list in Table \ref{tab:pb_lines}. We determined Pb abundances for three of the four GHOST stars and upper limits for the two Tull stars. The final fits to the Pb features are presented in Figure \ref{fig:pb}. We could not determine Pb abundances for the K23 sample as the wavelength range of the archival UVES spectra does not cover the Pb \textsc{i} $\lambda$4057.8 \AA~feature.

The Pb \textsc{i} $\lambda$4057.8~\AA~feature has significant departures from LTE, even among [Fe/H] \textgreater -1 stars, with departures as large as 0.30 dex \citep{Mashonkina2012}. To estimate approximate non-LTE corrections, we interpolate (for HD 216, extrapolate) Table 1 in \citet{Mashonkina2012}, estimating corrections of +0.26, +0.23, +0.26, +0.19, and +0.18 for UCAC4 278, UCAC4 179, UCAC4 445, HD 216, and TYC 1238, respectively.  We add these corrections to our [Pb/H] abundances before comparing to models.  Additionally, we add in quadrature an extra 0.10 uncertainty term to account for non-LTE correction uncertainties for [Pb/Fe] \citep{Mashonkina2012}. 

For all other elements besides Pb, we only report LTE abundances. Light elements are used only to confirm standard disk chemical evolution, so NLTE corrections are outside the scope of our work.  Furthermore, adopting NLTE corrections for elements not used in our model comparison (first-peak $s$-process elements) does not impact the outcomes of this work. NLTE corrections for \ion{Eu}{ii} $\lambda$6645 and the \ion{Y}{ii} lines used in this study are expected to be minimal near solar metallicities \citep{Storm2023, Storm2024, Mashonkina2026}. \ion{Ba}{ii} corrections are also minimal near solar metallicities, and even so, NLTE corrections depend not only on metallicity but also on absolute abundance, and we could not find corrections at these extreme levels of [$s$/Fe] enhancement. To our knowledge, the remaining elements also do not have reported NLTE corrections in the literature computed at these extreme abundances. All results should be interpreted with this caveat.

\subsubsection{Isotopic ratios}

We explore the isotopic implications of both the 
$^{13}\rm{C}(\alpha, n)^{16}\rm{O}$ and $^{22}\rm{Ne}(\alpha, n)^{25}\rm{Mg}$ neutron sources via both C and Mg isotopic analysis. 
The high resolution and wide wavelength coverage afforded by the GHOST and Tull spectrographs enable isotopic measurements of \cisotope within our sample via the 8000~\AA{} CN features, and \mgtwentyfour:\mgtwentyfive:\mgtwentysix from the MgH features around 5134~\AA{}. Appendix \ref{sec:isotope_methods} details the fitting procedure adopted for this work, with C and Mg isotopic ratios listed in Table \ref{tab:c12c13} and \ref{tab:mg_isotopes}, respectively.

While $^{12}$C is primarily synthesised through the triple-$\alpha$ reaction during helium burning, $^{13}$C is produced through proton captures on $^{12}$C during hydrogen burning via the CN cycle. Consequently, the $^{12}$C/$^{13}$C ratio is a sensitive diagnostic of stellar nucleosynthesis and internal mixing \citep[e.g.,][]{Karakas2014, Spite2006}. The solar isotopic ratio of $\sim 89$ decreases to $\sim 20$–$30$ following the first dredge-up on the red giant branch (RGB), and can approach the CN-cycle equilibrium value of $\sim 3$–$4$ through additional mixing \citep[e.g.,][]{Gratton2000, Charbonnel2010}. Conversely, third dredge-up during the AGB phase can substantially increase the surface $^{12}$C/$^{13}$C ratio through the enrichment of $^{12}$C synthesised during helium burning, with ratios exceeding 100 in some stellar models \citep[e.g.,][]{Karakas2014}.

\begin{figure*}
    \centering
    \includegraphics[width=\linewidth]{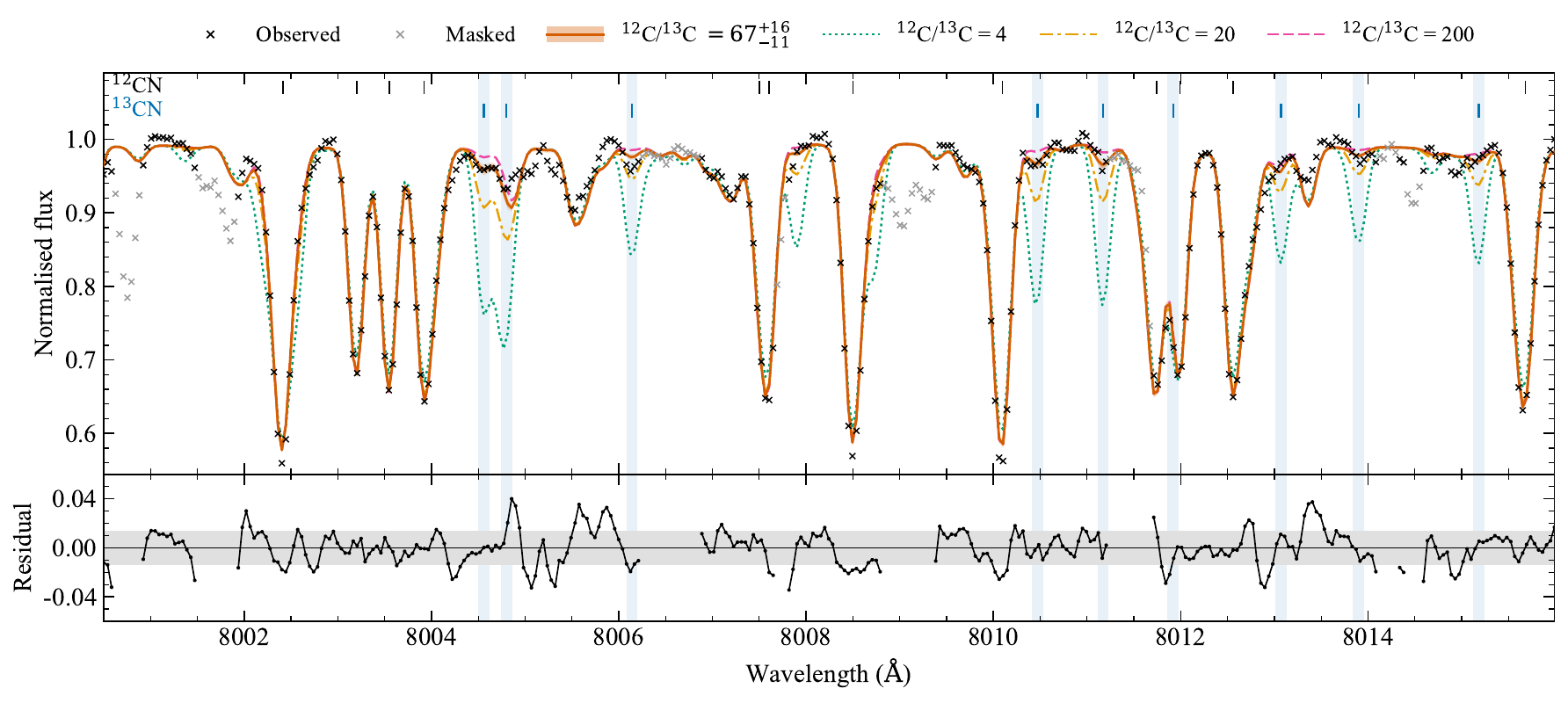}
    \caption{The 8000-8016\,\AA{} CN region for UCAC4 278-068826. Black crosses show the pixels used in the fit. Grey crosses show masked pixels, excluded because of known line-list deficiencies or removed by the iterative $3\sigma$ clipping. No pixels were masked for telluric absorption as none were deeper than our 1\% masking threshold within the fitted region for this star. The orange line and shading show the posterior median model and its 68\% interval, corresponding to $^{12}$C/$^{13}$C $= 66^{+16}_{-11}$. The other lines show the same model with only the isotopic ratio changed to $^{12}$C/$^{13}$C $= 4$ (dotted), 20 (dash-dotted) and 200 (dashed). Ticks at the top mark the strongest $^{12}$CN lines (black) and the $^{13}$CN features (blue, also shaded). The lower panel shows the residuals, excluding masked pixels, and the grey band indicates the $\pm1\sigma$ pixel noise.}
    \label{fig:c_iso}
\end{figure*}

Figure \ref{fig:c_iso} illustrates the CN region used to determine $^{12}$C/$^{13}$C ratios, highlighting the sensitivity of individual $^{13}$CN features to the isotopic composition. The synthetic spectra span a range of isotopic ratios representative of different nucleosynthetic and evolutionary conditions. A ratio of $^{12}$C/$^{13}$C = 4 approximates CN-cycle equilibrium, indicative of extensive hydrogen burning, while a ratio of 20 is characteristic of low-mass red giants following first dredge-up. At the opposite extreme, a ratio of 200 illustrates substantial $^{12}$C enrichment, as may arise from the dredge-up of helium-burning products in AGB stars. The comparison demonstrates how the strength of the $^{13}$CN features diminishes with increasing $^{12}$C/$^{13}$C, providing a diagnostic of the degree of CN processing and subsequent carbon enrichment. Appendix~\ref{sec:isotope_methods} outlines and discusses the specific fitting techniques.

In contrast to C isotopic ratios, the relative abundances of the three stable Mg isotopes (\mgtwentyfour, \mgtwentyfive, and \mgtwentysix) are largely unaffected by the internal evolution of low-mass stars, instead tracing the nucleosynthetic history of the material from which they formed. The dominant isotope, $^{24}$Mg, is produced primarily during hydrostatic C and Ne burning in massive stars, while the neutron-rich isotopes $^{25}$Mg and $^{26}$Mg are produced in both massive stars and intermediate-mass AGB stars. The production of neutron-rich Mg isotopes in massive stars is strongly metallicity-dependent, while AGB stars can provide additional enrichment through $^{22}$Ne($\alpha$,n)$^{25}$Mg and $^{22}$Ne($\alpha$,$\gamma$)$^{26}$Mg \citep[e.g.,][]{Kobayashi2011, Karakas2014, McKenzie2024MNRAS.534L..35M}. Consequently, Mg isotopic ratios provide a diagnostic of the chemical enrichment history of a stellar population, particularly the contributions from core-collapse supernovae and AGB stars. For comparison, the solar isotopic composition is $^{24}$Mg:$^{25}$Mg:$^{26}$Mg = 79:10:11, while Arcturus exhibits a similar ratio of approximately 82:9:9 \citep{McWilliam1988, Thygesen2016A&A...588A..66T}.

Mg isotopic ratios proved challenging to derive in our program stars because strong C$_2$ Swan bands contaminate the MgH features near 5130,\AA\ used for isotopic analysis. We employ the same spectral synthesis procedure developed and tested in \citet{McKenzie2024MNRAS.527.7940M, McKenzie2024MNRAS.534L..35M}, with the details of our fitting procedure described in Appendix~\ref{sec:isotope_methods}. Figure~\ref{fig:mg_iso} illustrates the sensitivity of the MgH absorption profiles to the relative contributions of the neutron-rich isotopes. The weaker $^{25}$MgH and $^{26}$MgH components produce subtle asymmetries in the dominant $^{24}$MgH features, allowing the isotopic composition to be constrained through detailed spectral synthesis. The surrounding molecular and atomic transitions are included in the synthesis to account for neighbouring blends, while the comparison with a heavy-isotope-enhanced mixture ($^{24}$Mg:$^{25}$Mg:$^{26}$Mg = 60:25:15) demonstrates the sensitivity of the fitted region to changes in the Mg isotopic composition.

\begin{figure}
    \centering
    \includegraphics[width=\linewidth]{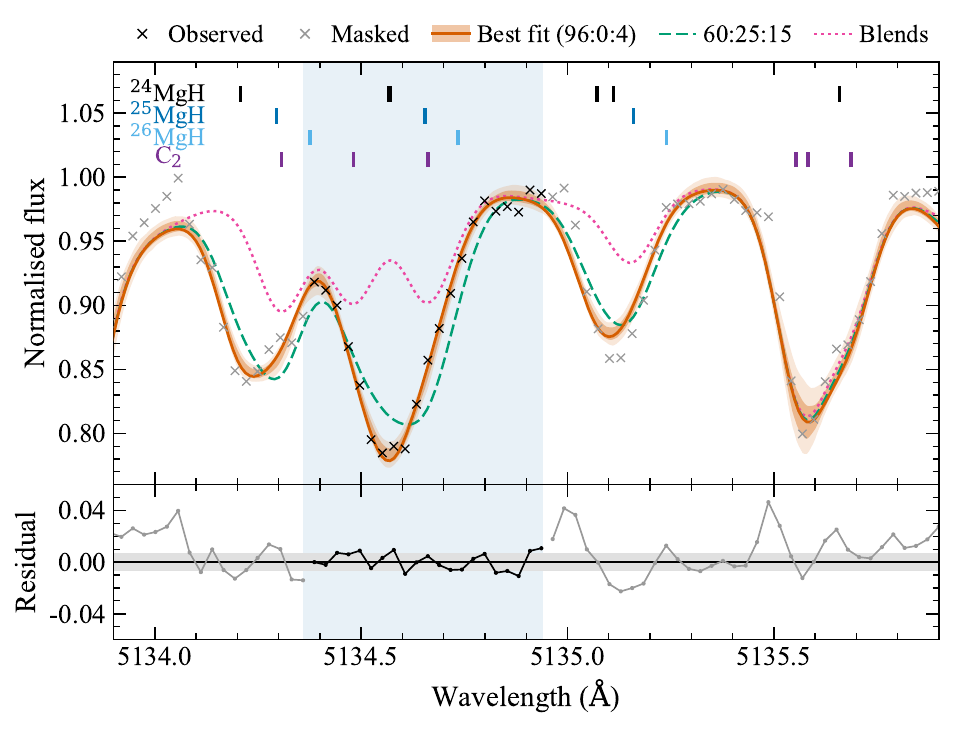}
    \caption{Fit to the R1 MgH feature at $5134.6$\,\AA\ in the GHOST spectrum of UCAC4 179-210559.
    Black crosses mark the pixels used in the fit (shaded blue region, 5134.36--5134.94\,\AA); grey crosses
    are masked. As in Fig. \ref{fig:c_iso}, the solid orange line is the best-fitting synthesis, with $^{24}$Mg:$^{25}$Mg:$^{26}$Mg$= 96{:}0{:}4$, and the shading shows the 68\% and 95\% intervals from posterior draws. For comparison, the dashed green line gives the same parameters with an illustrative heavy-isotope-enhanced mixture (60:25:15), of the kind expected from $^{22}\rm{Ne}(\alpha, n)^{25}\rm{Mg}$ and  $^{22}\rm{Ne}(\alpha, \gamma)^{26}\rm{Mg}$ burning. The dotted pink line is the best-fitting synthesis with all MgH transitions removed, showing the contribution of C$_2$, CN and atomic blends adopted from Table \ref{tab:abundances}. Ticks mark the strongest $^{24}$MgH, $^{25}$MgH, $^{26}$MgH and $^{12}$C$_2$ transitions. The window extends beyond the fitted region to show that the C$_2$ feature near 5135.6\,\AA\ is reproduced with the adopted C abundance and $^{12}$C/$^{13}$C $= 50$. The lower panel shows the observed minus best-fit residuals; the grey band marks $\pm1\sigma$ of the residuals in the fitted region. We note that we show the best-fit synthesis for this particular line, which is similar but not exactly the overall best-fit isotopic ratio of 94:3:3, as listed in Table \ref{tab:mg_isotopes}.}
    \label{fig:mg_iso}
\end{figure}

\subsection{AGB Model Selection and Fitting Procedure}
\label{sec:agb_models}

With abundances for our entire sample, we investigate whether enrichment from \iprocess nucleosynthesis can explain any of the abundance patterns by comparing to predictions from three sets of AGB nucleosynthesis models.  We start with K23, which computed a set of two 2 M$_\odot$, [Fe/H] $=$ -0.50 models using STAREVOL (see their Section 6 for a detailed description of their methods).  Both models are identical with the exception of their overshoot mixing prescriptions.  To our knowledge, the K23 \iprocess model is the only model of its kind near solar metallicity that enforces sufficient extra mixing to foster \iprocess conditions and efficient heavy-element production.  \citet{Choplin2024} also publishes a grid of AGB nucleosynthetic models with overshoot, but only two of their models experience proton ingestion at [Fe/H] $\sim$ -0.5, and due to weaker mixing parameters, neither are sufficiently enriched in second-peak neutron-capture elements to describe the extreme compositions of our sample.  As such, we can only test our abundance patterns against the single K23 \iprocess model.  Beyond the K23 models, we also compare to two widely used \sprocess AGB model sets: the Monash models \citep{Karakas2016, Karakas2018} and the FRUITY models \citep{Cristallo2011, Cristallo2015}. 

We detail our model fitting procedure in Appendix \ref{appendix:models}.  In short, we perform tests of each star's abundance pattern against models, considering among other things the effects of Galactic chemical evolution on the initial stellar abundances and the dilution of the mass transfer products (parameterised by variable $\delta$). For each model, we compute $\chi^2$ goodness-of-fit metrics and associated $p$-values, considering a model able to explain a star's abundance pattern if $p > 0.01$ \citep[e.g.,][]{Wall2012}.  We note that these $\chi^2$ and $p$-values assume uncorrelated Gaussian uncertainties and should be considered approximate \citep[see discussion in][]{Hampel2019}.

Our methodology intentionally makes it difficult for an \iprocess model to be favoured.  We are testing just one \iprocess model at fixed metallicity and mass to a large grid of \sprocess models that have greater pattern diversity and thus a higher likelihood of better fitting the observations.  Cases where the \iprocess model emerges as favoured could indicate that \iprocess nucleosynthesis is a plausible enrichment scenario.

\section{Results and Discussion}

\subsection{Abundance Behavior}

\subsubsection{Abundance Overview}
The final parameters and abundances for our sample are presented in Appendix Table~\ref{tab:abundances}.  We observe significant diversity in their abundance patterns, suggesting that a variety of processes can enrich stars to the extreme [Ba/Fe] and [Eu/Fe] (or [Sm/Fe] for the Tull stars) compositions on which we selected them.

We validate that all sample stars have [Ba/Fe] or [La/Fe] exceeding +1~dex, supporting the external enrichment scenario.  We confirm that three of the four GHOST stars selected by their extreme Eu in GALAH DR3 exhibit [Eu/Fe] \textgreater 1, but the fourth (UCAC4~179-210559) exhibits a lower [Eu/Fe] of $+0.59 \pm 0.11$, still suggestive of enrichment at these metallicities \citep[e.g.][]{GALAHDR3, gaiaeso2} but not at the magnitudes suggested by GALAH.  One of the two Tull stars selected for extreme [Sm/Fe], HD216, returns [Sm/Fe] of $+0.76 \pm 0.08$ for this element, not far below what GALAH reports (Table \ref{tab:observations}), but the other does not exhibit a significant enhancement.  For the entire sample, C, N, and O are enhanced beyond standard disk values to varying degrees, also indicative of external enrichment.  For the three stars where robust Pb abundances are possible, we determine extreme [Pb/Fe] values of $+1.67 \pm 0.19$, $+2.24 \pm 0.27$, and $+2.53 \pm 0.32$ for UCAC4~179-210559, UCAC4~445-009460, and UCAC4~278-068826, respectively.

Beyond elements significantly affected by AGB nucleosynthesis, most of the sample exhibits alpha, odd-Z, and iron-peak element abundances typical of disk star chemical evolution.  The only exception is UCAC4~265, which has a peculiar abundance pattern, with A(Li) near the primordial meteoritic value, extreme enhancements of K and Co exceeding 1~dex, and significant enhancements in V, Cr, and Mn.  At T$_{\rm eff}$ of 6587~K, this star may be experiencing atomic diffusion characteristic of chemically peculiar stars \citep[e.g.,][]{Tian2023_chempec}.

Our reanalysis of the two K23 stars returns parameters and abundances consistent with theirs within 0.10~dex for most elements, despite using different spectra and methods.  The only exception is [Eu/Fe], for which we determine $\sim$0.50~dex lower values, reducing their extreme Eu enhancements closer to predicted values from both the $s$ and $i$-processes (see Figure \ref{fig:eu} for our synthetic spectral fits to the Eu \textsc{ii} $\lambda$6645\AA~feature).

\subsubsection{Comparison to Literature}
\label{sec:lit_comparison}

\begin{figure*}
    \centering
    \includegraphics[width=\linewidth]{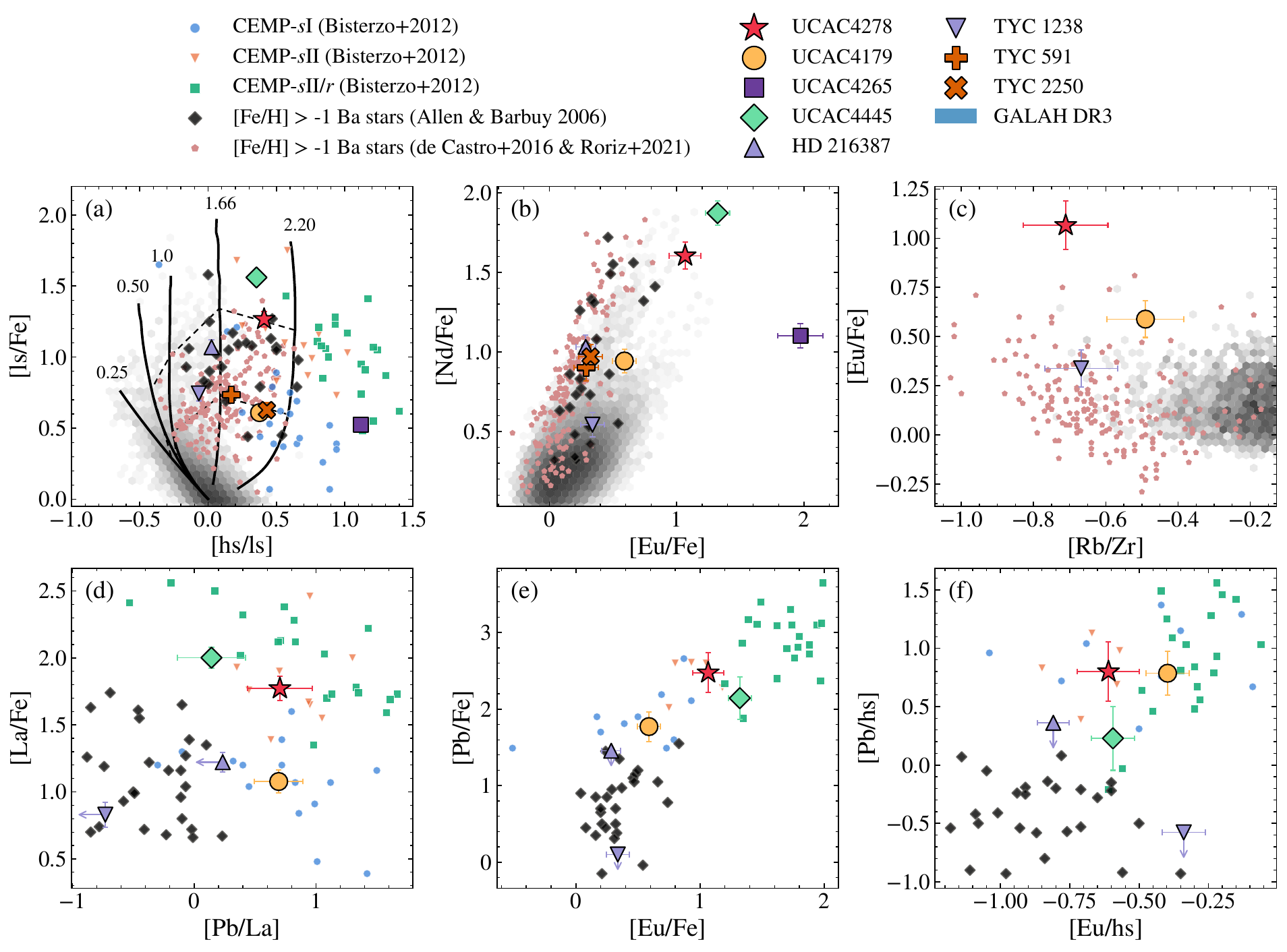}
    \caption{
    Our sample (colorful markers as indicated in the legend) in different abundance ratio planes, compared to the abundances for the $\rm [Fe/H] > -1$ barium stars from \protect\citet{Allen2006} (black diamonds) and \protect\citet{deCastro2016, Roriz2021_heavys} (gray pentagons), as well for the CEMP stars from \protect\citet{Bisterzo2012} of type $s$I (blue circles), $s$II (orange triangles), and $s$II/$r$ (green squares). \\
    \textit{Panel (a):} ``Hand Plot'' of \protect\citet{Busso2001}, which traces total enrichment ([ls/Fe], where dashed lines mark the fourth and eighth dredge up episodes) as a function of [hs/ls] (see Section~\ref{sec:lit_comparison}). The solid curves represent envelope models with constant $s$-process efficiency; the number above each curve records [N($^{13}$C)/N($^{56}$Fe)], tracing the neutron-to-seed ratio. Besides the chemically peculiar UCAC4 265, our sample does not stand out from other barium stars in this plane.\\
    \textit{Panel (b):} [Nd/Fe] as a function of [Eu/Fe]. Nd has greater contributions from the $s$-process whereas [Eu/Fe] is $r$-process dominated \protect\citep{Prantzos2020}.  One can draw an enrichment ``vector" anchored at the origin that traces $s$- vs. $r$-contributions, with lower [Eu/Fe] relative to [Nd/Fe] suggesting a stronger \sprocess vector.  Literature barium stars are dominated by the $s$-process vector, whereas our stars have stronger contributions from the $r$-process vector given their enhanced [Eu/Fe].\\
    \textit{Panel (c):} [Eu/Fe] as a function of [Rb/Zr], a tracer of neutron-density \protect\citep[e.g.,][]{Roriz2021_Rub}. Our stars exhibit negative [Rb/Zr] ratios as is typical of other barium stars, suggesting that the main neutron source is the $^{13}$C($\alpha$,n)$^{13}$O reaction.\\
    \textit{Panel (d):} [Pb/La] as a function of [La/Fe], which is similar to panel (a) but traces total neutron-exposure. Unlike in Panel (a), here, UCAC4 279, UCAC4 179, and UCAC4 445 stand out from typical barium stars and more closely resemble CEMP stars.\\
    \textit{Panel (e):} [Pb/Fe] as a function of [Eu/Fe]. The level of Pb and Eu enrichment in UCAC4 279, UCAC4 179, and UCAC4 445 mirrors that found in CEMP stars, while the remaining sample more closely resembles the compositions of typical barium stars.\\
    \textit{Panel (f):} [Pb/hs] as a function of [Eu/hs].  Once again, UCAC4 279, UCAC4 179, and UCAC4 445 mirror the behaviour of CEMP stars in this plane.\\
    }
    \label{fig:abundratios}
\end{figure*}

We use several diagnostic abundance ratios to compare our sample to the existing barium and CEMP star literature, including the $\sim$160 star [Fe/H] \textgreater -1 barium sample of \citet{deCastro2016, Roriz2021_heavys, Roriz2021_Rub}, the smaller sample of \citet{Allen2006} which reports Pb, and the CEMP sample of \citet{Bisterzo2012} (Figure~\ref{fig:abundratios}). 

First, we place these samples in the ``hand plot" of \citet{Busso2001}, which presents [ls/Fe] (here, ls is averaged over Sr, Y, and Zr) as a function of [hs/ls] (here, hs is averaged over La, Ce, and Nd, though some works also include Ba).  This plane is often used to trace \sprocess efficiency and the number of thermal pulses of the donor. Continuous curves in the hand plot represent AGB envelope models with different $^{13}$C-pocket efficiencies reflected in the [N($^{13}$C)/N($^{56}$Fe)] neutron-to-seed ratio. Dashed lines connect the 4th and 8th dredge up episodes, tracing stars' distributions along the thermally pulsing AGB. The underlying GALAH DR3 population is predominantly concentrated at negative [hs/ls], in contrast to our sample, which favours positive values. This suggests preferential enrichment of the second $s$-process peak relative to the first in our barium stars compared with the general Galactic population. Besides the chemically peculiar UCAC4~265-014129, our sample does not stand out significantly against other barium and CEMP stars in this plane.

Next, we compare [Nd/Fe] and [Eu/Fe] abundances, where [Nd/Fe] has approximately equal contributions from $s-$ and $r-$ process nucleosynthesis, and [Eu/Fe] is predominantly produced by the \rprocess in the solar neighbourhood \citep{Prantzos2020}. The underlying GALAH DR3 population exhibits an apparent bifurcation towards higher [Eu/Fe], separating into two sequences with different [Nd/Eu] ratios. We interpret these as two primary ``enrichment vectors,'' one of which traces \sprocess enrichment (favouring Nd production over Eu) and the other possibly \rprocess enrichment or standard disk evolution.  Our sample predominantly occupies the Nd-rich sequence, alongside other barium stars, though UCAC4~278-068826, UCAC4~179-210559, and UCAC4~445-009460 lie closer to the gap in the bifurcation, suggesting greater [Eu/Fe] at fixed [Nd/Fe]. The origin of the apparent bifurcation in the broader GALAH population remains unclear. UCAC4~265-014129 once again stands out, here for its extreme [Eu/Fe].

[Rb/Zr] is a commonly used diagnostic of neutron density during AGB nucleosynthesis \citep[e.g.,][]{GH2006Sci...314.1751G, vanRaai2012A&A...540A..44V, PM2017A&A...606A..20P}. The production of Rb is sensitive to branching at $^{85}$Kr, with higher neutron densities favouring the accumulation of $^{87}$Rb relative to neighbouring first-peak elements such as Zr. Consequently, negative [Rb/Zr] ratios are generally associated with the low neutron densities of the $^{13}$C($\alpha$,n)$^{16}$O neutron source, whereas positive ratios can indicate activation of $^{22}$Ne($\alpha$,n)$^{25}$Mg in more massive AGB stars. Our sample exhibits negative [Rb/Zr], suggesting that the $^{13}$C($\alpha$,n)$^{16}$O is the primary neutron source. We note that while GALAH reports Rb abundances, the 7800\AA{} line is heavily blended with Si and can occasionally be poorly fit by the abundance pipeline \cite{Buder2026arXiv261025583B}.  

Next, we compare our sample to barium and CEMP stars in the [La/Fe] vs. [Pb/La] plane, which is similar in nature to the ``Hand Plot" of Panel (a) but instead traces neutron exposure through the second-to-third peaks.   This diagnostic ratio is particularly useful for our Pb-rich stars.  At high neutron exposures, the \sprocess flows passes through the first two peaks and accumulates at Pb, with [hs/ls] reaching an equilibrium at extreme exposures while [Pb/hs] continues to rise \citep[e.g.][]{Busso2001, Hampel2019}. In our Pb-rich stars, [hs/ls] is moderately enhanced ($+0.3$ to $+0.4$) whereas [Pb/hs] reaches $+0.3$ to $+0.8$, suggesting that [hs/ls] underestimates the neutron exposure experienced by these stars and that [Pb/hs] is more informative.

We find that literature stars cluster into groups in the [La/Fe] vs. [Pb/La] plane.  The [Fe/H] \textgreater -1 barium star sample of \citet{Allen2006} exhibits solar-to-sub-solar [Pb/La].  The CEMP-$s$I stars of \citet{Bisterzo2012} occupy the high-[Pb/La], low-[La/Fe] (where low is relative--these are still enriched to $\sim$+1 dex) quadrant.  Finally, the CEMP-$s$II and -$s$II/r stars lie in the high-[Pb/La], high-[La/Fe] region of the plane. HD 216 and TYC 1238, the two stars for which we only have Pb upper limits, behave much like the barium stars of \citet{Allen2006}, exhibiting roughly solar or subsolar [Pb/La].  The remaining three stars for which we have Pb--UCAC4~278-068826, UCAC4~179-210559, and UCAC4~445-009460-- have extreme [Pb/La] that resemble CEMP stars more closely; UCAC4~179-210559 being CEMP-$s$I-like and the others CEMP-$s$II and -$s$II/r-like. This behaviour could suggest that the extreme neutron-exposures associated with CEMP stars at lower metallicities may also be responsible for the extreme Pb compositions of these stars \citep[e.g.,][]{Bisterzo2012, Hampel2019}.

Finally, we study the relationship between Eu and Pb in our sample. UCAC4~278-068826, UCAC4~179-210559, and UCAC4~445-009460 stand out in the [Pb/Fe] vs. [Eu/Fe] plane due to their extreme enrichment in both elements that set them apart from other metal-rich barium stars.  Once again, their abundance pattern mimics those of CEMP-$s$I/$s$II/$r$ stars, and to our knowledge, these stars represent the most Pb-rich, near solar metallicity barium stars known.  Similarly, in both Panel (b) and (e), UCAC4~278-068826 and UCAC4~445-009460 exceed the [Eu/Fe] of all metal-rich barium stars to which we compare here.  In Section \ref{sec:comparisontomodels}, we compare these abundance patterns to predictions from AGB models to seek an explanation for their extreme Eu and Pb compositions.

Not shown in Figure \ref{fig:abundratios} is our comparison to the [W/Fe] compositions of metal-rich barium stars from \citet{Roriz2021_W}, which discusses how extreme W enhancements could be indicative of enhanced neutron densities characteristic of the \iprocess.  We find that our sample exhibits [W/hs] similar to the average metal-rich barium star in their sample.

\subsection{Isotopic Behavior}
We make a concerted effort to derive isotopic ratios for both C and Mg in our sample of stars, since each provides valuable constraints on neutron production through the two neutron-source reactions $^{13}$C($\alpha$,n)$^{16}$O and $^{22}$Ne($\alpha$,n)$^{25}$Mg. We obtain a well-constrained $^{12}$C/$^{13}$C ratio of $66^{+16}_{-11}$ for UCAC4~278-068826, while TYC~1238 and UCAC4~179-210559 have lower limits of 19 and 18, respectively (Table \ref{tab:c12c13}). Reliable isotopic constraints could not be obtained for the remaining stars owing to low SNR, weak CN features, or telluric contamination. The measured ratio and lower limits are inconsistent with CN-equilibrium values ($^{12}$C/$^{13}$C $\sim 3$–4), with UCAC4 278-068826 exhibiting a particularly high ratio indicative of substantial $^{12}$C enrichment.

UCAC4 278-068826 lies on the lower red giant branch (Figure \ref{fig:sample}), where first dredge-up is expected to reduce the surface $^{12}$C/$^{13}$C ratio. Its measured value is nevertheless higher than the ratios of 20–30 typically observed in normal giants following first dredge-up \cite[e.g., see][]{Karakas2014}. This suggests that its surface carbon remains dominated by $^{12}$C-rich material, consistent with enrichment through mass transfer from an evolved companion. The extent to which first dredge-up modifies the isotopic composition depends on the amount and composition of the accreted material, making the elevated ratio a useful constraint on its subsequent mixing history \cite[e.g.,][]{Stancliffe2008MNRAS.389.1828S, Stancliffe2009MNRAS.396.2313S}. TYC~1238-203-1  and UCAC4~179-210559 occupy positions near the base of the red giant branch and on the subgiant branch, respectively, where first dredge-up may be underway, but substantial extra mixing associated with the RGB bump is not expected.

We derive Mg isotope ratios for two stars where the MgH features are visible but not too strongly contaminated by C$_2$ features. For TYC~2250-1047-1 we find that a Mg isotopic ratio of 83:8:9 provides the best fit to the spectrum. This agrees with isotopic ratios from Arcturus \citep{McWilliam1988, Thygesen2016A&A...588A..66T} and aligns with expectations from Galactic Chemical Evolution \citep{Kobayashi2011, Kobayashi2020}. UCAC4~179-210559, however, paints a very different story. Its isotopic ratio of 94:3:3 is more similar to stars with [Fe/H] abundances at least one dex lower, at [Fe/H] $\approx-1.5$ \cite[e.g., see][]{Gay2000ApJ...533..260G, Yong2003ApJ...599.1357Y, McKenzie2024MNRAS.534L..35M}. Fig. \ref{fig:mg_iso} also demonstrates that the asymmetries in MgH lines are not driven by MgH itself, but by overlying blends. All lines used in the analysis (listed in Table \ref{tab:mg_isotopes}) point to a very low heavy Mg isotopic mixture for this star.

The production of the neutron-rich isotopes $^{25}$Mg and $^{26}$Mg is strongly metallicity-dependent, with their relative abundances generally increasing with [Fe/H] \citep{Yong2003ApJ...599.1357Y}. At low metallicity, massive stars predominantly produce the primary isotope $^{24}$Mg, while the production of the neutron-rich isotopes $^{25}$Mg and $^{26}$Mg is suppressed by the reduced availability of neutron-rich seed nuclei. The combination of enhanced [Mg/Fe] and low neutron-rich Mg isotopic fractions may therefore reflect an initially $^{24}$Mg-rich composition inherited from the star’s natal environment.

The contrast between the Mg isotopic compositions of our two stars is particularly interesting given their similar metallicities and [Mg/Fe] enhancements ($+0.36$ and $+0.30$ for UCAC4179-210559 and TYC2250-1047-1, respectively). While the neutron-capture enrichment in these stars may be associated with different AGB nucleosynthetic conditions, neither conventional low-mass AGB nucleosynthesis nor PIEs necessarily predict a depletion of both neutron-rich Mg isotopes. In particular, although the $^{13}$C($\alpha$,n)$^{16}$O neutron source can produce substantial neutron-capture enrichment without significant activation of $^{22}$Ne($\alpha$,n)$^{25}$Mg, this would not readily explain the unusually low $^{25}$Mg and $^{26}$Mg fractions in UCAC4~179-210559. Its isotopic composition may therefore reflect an unusual natal chemical enrichment history, subsequently preserved during AGB mass transfer. This anomalous Mg isotopic abundance warrants re-observations with a higher-resolution spectrograph (R $\gtrsim80,000$) to better resolve the subtle isotopic splitting. We will continue to investigate this phenomenon in future work.

\subsection{Comparison to Models: The Most Promising Candidates for \iprocess Enrichment}\label{sec:comparisontomodels}
\begin{figure*}
    \centering
    \includegraphics[width=.9\linewidth]{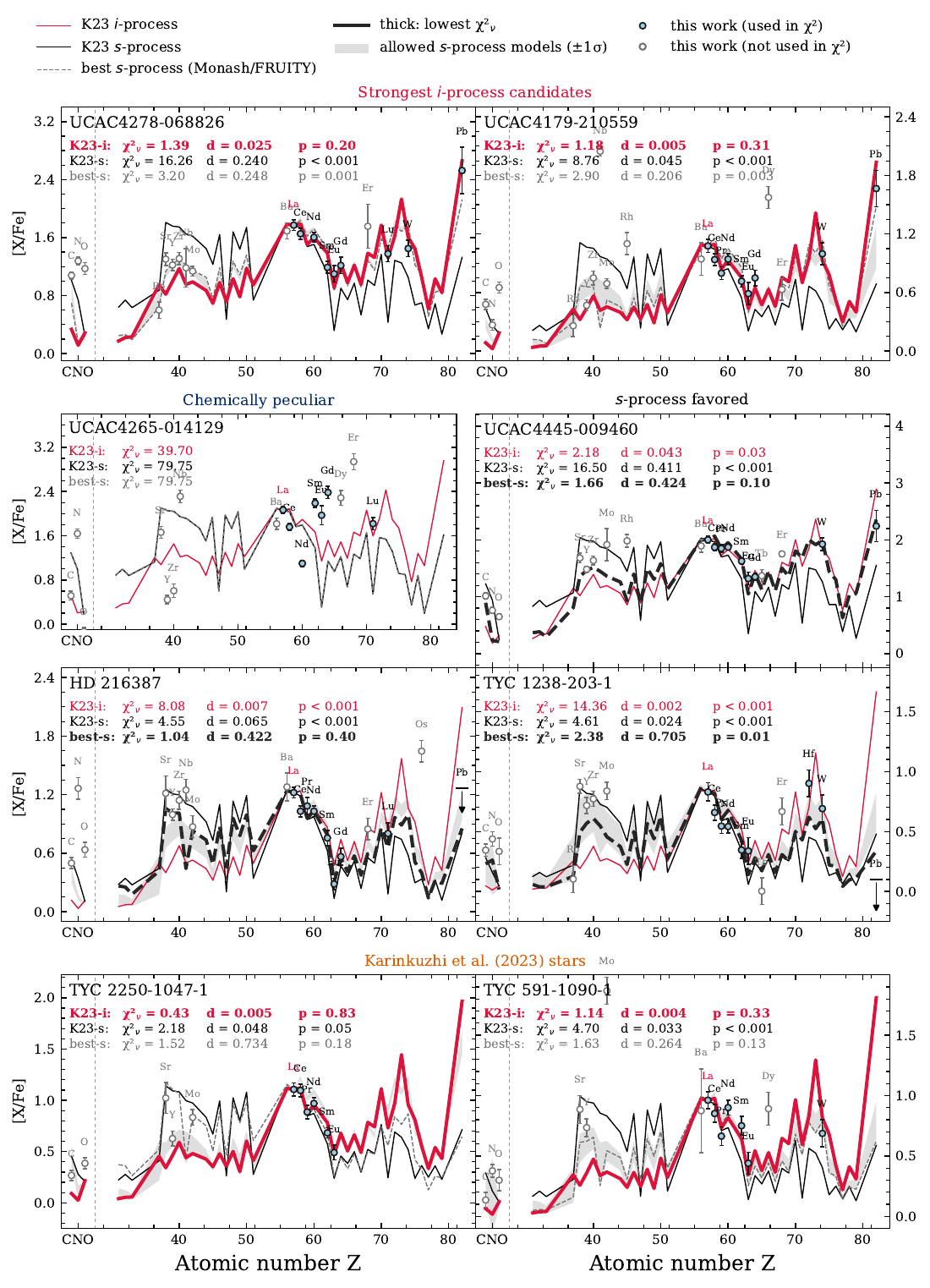}
    \caption{[X/Fe] abundances for C, N, O, and the neutron-capture elements compared to diluted K23 $i$- (solid red) and $s$-process (solid black) AGB models as well as the best-fitting diluted Monash or FRUITY $s$-process model (dashed). $\chi^2$ normalised to the number of elements in the fit (where blue markers represent fitted elements), $\delta$, and the associated $p$-value for each model are included in each panel. The 1$\sigma$ range of plausible $s$-process models is shaded grey. Two new promising $i$-process-enriched candidates emerge (top row).}
    \label{fig:model_comparison}
\end{figure*}

In Figure \ref{fig:model_comparison}, we present the comparison of the abundances of our sample to the diluted K23 $s$- and $i$-process models as well as the best-fit plausible Monash or FRUITY model (Section~\ref{sec:agb_models} and Appendix~\ref{appendix:models}). The $\chi^2$, $\delta$, and detailed parameters associated with each model are presented in Appendix Table \ref{tab:models}.  

Ultimately, two stars emerge as promising new candidates of \iprocess enrichment: UCAC4~278-068826 and UCAC4~179-210559.  These two stars have abundance patterns and absolute enrichment levels that are well-matched to the K23 \iprocess model and not well-explained by any of the tested \sprocess models ($p \leq 0.003$).  UCAC4~278-068826 returns a reduced $\chi^2$ ($\chi^2_\nu$) value of 1.39 ($p$=0.20) when compared to the diluted K23-$i$ model yet a $\chi^2_\nu$ of 3.20 ($p$=0.001) against the best-fit Monash model, a 1.81 difference in $\chi^2_\nu$.  Similarly, UCAC4~179-210559 returns a $\chi^2_\nu$ value of 1.18 ($p$=0.31) for the K23-$i$ model, while the comparison to the best-fit FRUITY model yields a $\chi^2_\nu$ of 2.90 ($p=0.003$), a 1.72 improvement in the $\chi^2_\nu$ value when adopting the \iprocess model.  Among the measured elements, Pb plays the most important role in distinguishing $i$- from $s$- enrichment.  UCAC4~179-210559 and UCAC4~278-068826 have some of the highest [Pb/Fe] abundances of the sample, approaching (for the former) and even exceeding (for the latter) 100$\times$ the Pb fraction of the Sun.

In addition to identifying UCAC4~278-068826 and UCAC4~179-210559  as stars well-explained by \iprocess enrichment, we validate that the two K23 stars, TYC~2250-1047-1 and TYC~591-1090-1, also favor the \iprocess models over the best-fit $s$-, with their fits to the K23-$i$ model returning $\chi^2_\nu$ values of 0.43 ($p$=0.83) and 1.14 ($p$=0.33), respectively.  For the former, this is a 1.09 improvement to that of the best-fit $s$- model ($\chi^2_\nu$ = 1.52); however, we cannot rule out the $s$- model ($p$=0.18).  For the latter star, the $\chi^2_\nu$ improvement is 0.50 from the best-fit $s$ model ($\chi^2_\nu$ = 1.63, $p$=0.13), and the $p$-values suggest that both the $s$- and $i$-process models can explain its abundance pattern.  Unfortunately, the archival UVES spectra do not cover the Pb \textsc{i} $\lambda$4057.8 \AA~feature. Future Pb follow-up could be able to distinguish between these two enrichment scenarios.

Three of the remaining stars favour the \sprocess models.  UCAC4~445-009460's abundances marginally favour a Monash $s$-process model over the K23 $i$-process model, though the \iprocess model cannot be fully ruled out ($p$=0.03).  Its non-LTE-corrected [Pb/Fe] of +2.24 $\pm$ 0.27 is $\sim$ 0.75 dex below what is predicted by the diluted \iprocess model.  Removing Pb from the fit shifts the preference to marginally favour the \sprocess model.  Overall, the abundance pattern is not very well fit by any model. The two Tull stars, HD 216 and TYC 1238, both strongly favour the best-fit \sprocess model, with the former well-described ($\chi^2$ = 1.04, p = 0.40) by a 2 $\rm M_\odot$ FRUITY model and the latter favouring a 1.5 $\rm M_\odot$ Monash model, though the fit is poor ($\chi^2$ = 2.38, p = 0.01).  The final star in the sample, UCAC4~265-014129, unsurprisingly does not fit to any tested AGB model given its unusual abundance pattern that may be indicative of atomic diffusion.

\section{Conclusions}
In this work, we analyse Gemini/GHOST, McDonald/Tull, and VLT/UVES spectra of eight $\rm [Fe/H] > -0.7$ barium stars to search for signs of \iprocess nucleosynthesis in their abundance patterns.  Six of these stars were selected from GALAH DR3 by their extreme [Ba/Fe] (\textgreater +1) and either extreme [Eu/Fe] (\textgreater +1) or [Sm/Fe] ($\sim$+1) and followed up for the first time here.  The remaining two are literature candidates.  We determine fundamental parameters, \cisotope, Mg isotope ratios, and abundances in up to 42 elements from Li through Pb and compare their chemical compositions to published $s$- and $i$-process AGB nucleosynthesis models. Only one tested $i$-process model achieves the extreme heavy-element enrichment observed in these stars, highlighting the limited theoretical coverage of this regime.

Three stars in our sample exhibit extreme Pb enhancements (+1.79$\pm$0.19 \textless [Pb/Fe] \textless +2.50$\pm$0.32 dex), possibly making them the most Pb-rich, [Fe/H] \textgreater -1 barium stars known. Two of these stars, UCAC4278-068826 and UCAC4179-210559, provide compelling evidence for $i$-process nucleosynthesis near solar metallicity. Their abundance patterns are well reproduced by the $i$-process model but are not well explained by any of the tested $s$-process models. Crucially, their extreme Pb enrichments, approaching or exceeding 100$\times$ solar, cannot be reproduced by the tested $s$-process models.  We revisit the existing literature \iprocess candidates and confirm that they favour the \iprocess model, but for these, \sprocess models cannot be ruled out --- in part because we could not determine [Pb/Fe] for these stars, which could be helpful for distinguishing between nucleosynthetic signatures.  The rest of the sample is consistent with \sprocess enrichment, highlighting that independent of $s$- versus $i$-, AGB stars can be efficient producers of Eu and other elements traditionally associated with the $r$-process. The only exception is one star that standard stellar nucleosynthesis cannot explain and is likely experiencing atomic diffusion.

Our work presents compelling evidence for the operation of the \iprocess near solar metallicities and confirms Pb as an effective discriminator of $s$- vs. $i$-process signatures.  These results raise questions about the origins of classically $r$-process elements such as Eu in the Galactic disk by demonstrating that AGB stars can produce this element in significant quantities.  Finally, our work highlights the importance of developing a larger grid of metal-rich \iprocess AGB models that span a range of mixing prescriptions, metallicities, and masses.  Such models would enable broader searches for \iprocess signatures among barium stars and allow its contributions to Galactic chemical evolution to be quantified.

\section*{Acknowledgements}

The authors are grateful to Vini Placco for assisting us with planning our observations and early interpretation of our results; to Drisya Karinkuzhi for her insights on the K23 sample, comparison to models, and expectations for \iprocess nucleosynthesis; to Marco Pignatari for insightful conversations about barium star characterization during the AstroNuc2026 workshop in Tuscon and for introducing us to the classic \citep{Busso2001} \sprocess ``hand plot,'' and to Andrew McWilliam for providing us with our CN line list, and for his helpful advice on stellar spectroscopy and nucleosynthesis. We are also grateful to the Gemini/GHOST data collection and analysis team for assisting us with collecting and reducing our observations and to the McDonald Observatory staff, including observing support and those running the Astronomer's lodge, for supporting our collection of the Tull spectra.

CM is supported by the NSF Astronomy and Astrophysics Fellowship award number AST-2401638. EM and AW were also supported by CM's NSF award (AST-2401638).  Support for this work was provided by NASA through the NASA Hubble Fellowship grant \#HST-HF2-51560 awarded by the Space Telescope Science Institute, which is operated by the Association of Universities for Research in Astronomy, Inc., for NASA, under contract NAS5-26555.

This research has made use of The Spectra Pointer database, operated at Georgia State University, Atlanta, Georgia.  This work is based on observations made with ESO Telescopes at the La Silla Paranal Observatory under programme ID 113.26NB.001. This work is based on observations obtained acquired through the Gemini Observatory Archive at NSF NOIRLab and processed using DRAGONS (Data Reduction for Astronomy from Gemini Observatory North and South) at the international Gemini Observatory, a program of NSF NOIRLab, which is managed by the Association of Universities for Research in Astronomy (AURA) under a cooperative agreement with the U.S. National Science Foundation on behalf of the Gemini Observatory partnership: the U.S. National Science Foundation (United States), National Research Council (Canada), Agencia Nacional de Investigación y Desarrollo (Chile), Ministerio de Ciencia, Tecnología e Innovación (Argentina), Ministério da Ciência, Tecnologia, Inovações e Comunicações (Brazil), and Korea Astronomy and Space Science Institute (Republic of Korea). 

This work used the Third Data Release of the GALAH Survey (Buder et al. 2021). The GALAH Survey is based on data acquired through the Australian Astronomical Observatory, under programs: A/2013B/13 (The GALAH pilot survey); A/2014A/25, A/2015A/19, A2017A/18 (The GALAH survey phase 1); A2018A/18 (Open clusters with HERMES); A2019A/1 (Hierarchical star formation in Ori OB1); A2019A/15 (The GALAH survey phase 2); A/2015B/19, A/2016A/22, A/2016B/10, A/2017B/16, A/2018B/15 (The HERMES-TESS program); and A/2015A/3, A/2015B/1, A/2015B/19, A/2016A/22, A/2016B/12, A/2017A/14 (The HERMES K2-follow-up program). We acknowledge the traditional owners of the land on which the AAT stands, the Gamilaraay people, and pay our respects to elders past and present. This paper includes data that has been provided by AAO Data Central (datacentral.org.au).

\section*{Data Availability}

The Gemini/GHOST spectra are publicly available in the Gemini Observatory Archive (Program GS-2026A-FT-104). The McDonald/Tull spectra are available by request to the authors. The UVES spectra are available in the ESO Science Archive (Program 113.26NB.001).



\bibliographystyle{mnras}
\bibliography{bibli} 




\appendix

\section{Observations and Data Reduction}\label{appendix:obs}
\subsection{Gemini/GHOST}
Spectra of four stars in our sample were collected with Gemini South's GHOST spectrograph, an optical (347--1060 nm), high-resolution (R $\sim$ 50,000--76,000), high efficiency instrument designed to maximise blue throughput \citep{Kalari2024}.  We used standard resolution mode (R$\sim$ 50,000) with 1$\times$2 binning (spectral$\times$spatial) for our observations.  Exposure times were estimated using the Gemini Integration Calculator and are listed in Table \ref{tab:observations} alongside our estimates of the resulting signal-to-noise (SNR) near 4500\AA in our spectra.

Spectra were reduced through the DRAGONS-integrated GHOST data reduction pipeline \citep{Placco2024, DRAGONS}, which outputs fully reduced, extracted, and combined one-dimensional spectra for both the blue and red GHOST arms, respectively.  These spectra were then continuum-normalised with \texttt{SUPPNET} \citep{SUPPNET} before being concatenated into a single spectrum, discarding the blue arm in favour of the higher SNR red arm in the overlapping region (5250–5450\AA).  These spectra are publicly available on the Gemini Observatory Archive under program GS-2026A-FT-104.  Radial velocities were determined for each spectrum using \texttt{iSpec} \citep{ispec}, and spectra were shifted to rest air wavelengths to prepare for analysis.

\subsection{McDonald/Tull}
Spectra of two stars were collected with the Tull spectrograph \citep[][]{Tull1995}, an optical (3500 \AA~ \textless~$\lambda$ \textless~10,000 \AA) high-resolution (R$\sim$60,000, Slit \#4) echelle spectrograph on the 2.7m Harlan J. Smith Telescope at McDonald Observatory.  Spectra were reduced using the Tull Coud\'{e} Spectrograph Data Reduction Pipeline (\texttt{TSDRP}\footnote{\url{https://github.com/grzeimann/TSDRP}}), which performs standard reduction steps. Additionally, \texttt{TSDRP} provides deblazing, continuum normalisation, and order combination to produce a single, fully processed spectrum.  Radial velocities were determined for each spectrum using \texttt{iSpec}, and spectra were shifted to rest air wavelengths.  Finally, sub-exposures of the same object were coadded by spline interpolating fluxes onto a shared wavelength grid and summing them together, weighing each flux array by the SNR of the subexposure.

\subsection{VLT/UVES}
K23 observed their \iprocess candidates with UVES using the 0.5" slit in both arms, giving R $\sim$ 74,450 in the red arm (472–683 nm) and R $\sim$ 65,030 in the blue arm (302–391 nm).  Spectra were found through the Spectra Pointer database,\footnote{\url{https://doi.org/10.5281/zenodo.22698835}} downloaded from the ESO archive, and then continuum-normalised with a cubic smoothing spline fitted to the 92nd-percentile flux in 8\AA~bins, with three iterations of $2.5\sigma$ clipping. Finally, radial velocity correction was performed with \texttt{iSpec}.

\section{Parameters and Abundances}\label{appendix:abunds}
We measure fundamental parameters (T$_{\rm eff}$, log(g), [M/H], and microturbulence, v$_{mic}$) and most abundances with version 72 of Brussels Automatic Code for Characterizing High accUracy Spectra (\texttt{BACCHUS}\footnote{\url{https://www.astro.ulb.ac.be/pmwiki/Spectro/Bacchus}}, \citealp{BACCHUS}) which is described in detail in several works \citep[e.g.,][]{Masseron2016, Sinha2026, Myers2026}. \texttt{BACCHUS} is an automated tool for analysing high-resolution spectra.  It uses LTE Turbospectrum \citep[e.g.,][]{Plez2012} for radiative transfer and determines line-by-line abundances using on-the-fly spectral synthesis.  We adopted MARCS model atmospheres and assumed Solar abundances from \citet{Grevesse2007}.  

Stellar parameters are determined with Fe ionisation-excitation balance, and line-by-line abundances are determined by comparing observations to synthetic spectra using four methods (see the cited literature which describes these methods thoroughly).  Each fitting method returns a flag that reflects the fidelity of the output abundance for the line.  We only consider abundances of elements where all four \texttt{BACCHUS} flags return 1, indicating a reliable fit to the line. We present our line selection in Table \ref{tab:lineselection}.

Above 4200 \AA, we adopted atomic and molecular data from the Gaia-ESO line lists \citep{Heiter2021}, which include hyperfine splitting for numerous species (Sc{\sc i}, V{\sc i}, Mn{\sc i}, Co{\sc i}, Cu{\sc i}, Ba{\sc ii}, La{\sc ii}, Pr{\sc ii}, Nd{\sc ii}, Sm{\sc ii} and Eu{\sc ii}), and isotopic splitting for Li, Cu, Ba and Nd.  We supplement with additional molecular data from numerous sources, all of which come pre-loaded with \texttt{BACCHUS}: CH from \citealt{Masseron2014}, C2, CN, OH, and MgH from T. Masseron, private communication, SiH from \citealt{Kurucz1992}, and TiO, FeH, and ZrO from B. Pelz, private communication.  

Below 4200\AA, we use an atomic line list built from the Vienna Atomic Line Database (VALD3; \citealt{Ryabchikova2015}). Hyperfine splitting is included for numerous species (Ba{\sc ii}, La{\sc ii}, Pr{\sc ii}, Lu{\sc ii}, Eu{\sc ii}, Tb{\sc ii}, Ho{\sc ii}, Yb{\sc ii}, Os{\sc i}, and Pb{\sc i}).  We supplement with molecular transition information from \texttt{linemake} \citep{linemake}. The only exception is the $\lambda$4050 -- 4070\AA~window for which we use \texttt{linemake} to make a dedicated line list.  That line list is included in Table \ref{tab:pb_lines}.

\begin{table*}
\centering
\caption{Our line selection, presenting every line used in the final abundance of at least one star. For each star, we only consider lines that are unflagged in all four \texttt{BACCHUS} methods (Section \ref{appendix:abunds}).}
\label{tab:lineselection}
\scriptsize
\setlength{\tabcolsep}{3pt}
\renewcommand{\arraystretch}{0.95}
\begin{tabular}{lcp{0.88\textwidth}}
\hline
Element & $N$ & Wavelengths (\AA) \\
\hline
Li & 1 & 6707.8 \\
C & 65 & 4216.8, 4236.7, 4242.9, 4249.2, 4255.4, 4261.6, 4273.8, 4275.3, 4279.9, 4282.0, 4284.2, 4285.9, 4288.8, 4291.1, 4292.1, 4293.1, 4295.1, 4297.3, 4303.8, 4304.4, 4306.1, 4306.8, 4323.9, 4347.9, 4359.5, 4366.6, 4371.1, 4374.2, 4377.2, 4380.8, 4387.1, 5029.9, 5037.7, 5045.2, 5052.2, 5052.6, 5056.1, 5059.8, 5063.1, 5066.7, 5070.0, 5073.4, 5086.2, 5089.4, 5092.3, 5103.8, 5113.1, 5114.3, 5135.6, 5141.2, 5144.6, 5150.6, 5153.1, 5155.5, 5157.6, 5159.4, 5161.1, 5165.1, 5380.3, 5540.2, 6587.0, 6587.6, 7111.5, 7113.2, 8727.0 \\
N & 68 & 7901.9, 7927.9, 7933.6, 7948.8, 7950.4, 7951.7, 7956.7, 7957.0, 7962.6, 7963.2, 7968.5, 7977.2, 7981.1, 7984.6, 7987.9, 7992.3, 8000.2, 8002.4, 8010.1, 8017.1, 8018.1, 8025.9, 8043.9, 8044.4, 8053.1, 8060.2, 8062.6, 8064.1, 8067.3, 8072.4, 8074.4, 8082.6, 8093.0, 8095.9, 8103.8, 8107.2, 8114.9, 8118.8, 8126.2, 8130.6, 8136.2, 8138.0, 8150.1, 8167.6, 8168.1, 8169.2, 8174.1, 8175.1, 8181.3, 8188.1, 8203.2, 8218.5, 8237.5, 8246.8, 8247.3, 8257.5, 8266.5, 8273.5, 8278.9, 8289.0, 8298.5, 8301.5, 8314.8, 8320.9, 8333.4, 8337.9, 8383.3, 8419.9 \\
O & 6 & 6300.3, 6363.8, 6366.0, 7771.9, 7774.2, 7775.4 \\
Na & 16 & 4276.8, 4393.3, 4494.2, 4497.7, 4664.8, 4668.6, 4751.8, 4978.6, 4982.8, 5148.8, 5682.6, 5688.2, 6154.2, 6160.8, 8183.2, 8194.8 \\
Mg & 12 & 4571.1, 4703.0, 4730.0, 5528.4, 5711.1, 6318.7, 6319.2, 6319.5, 8712.7, 8717.8, 8736.0, 8923.6 \\
Al & 3 & 5557.1, 6696.0, 6698.7 \\
Si & 15 & 5645.6, 5665.6, 5684.5, 5690.4, 5701.1, 5708.4, 5772.1, 5793.1, 5948.5, 6347.1, 6371.4, 6741.6, 8892.7, 8899.2, 8949.1 \\
S & 7 & 6743.4, 6743.5, 6743.6, 6748.6, 6748.7, 6748.8, 6757.2 \\
K & 1 & 7699.0 \\
Ca & 25 & 5001.5, 5260.4, 5261.7, 5349.5, 5513.0, 5582.0, 5588.7, 5590.1, 5857.4, 5867.6, 6102.7, 6122.2, 6156.0, 6162.2, 6166.4, 6169.0, 6169.6, 6439.1, 6455.6, 6456.9, 6471.7, 6493.8, 6499.6, 8912.1, 8927.4 \\
Sc & 13 & 5031.0, 5083.7, 5356.1, 5484.6, 5520.5, 5526.8, 5641.0, 5657.9, 5667.1, 5684.2, 5686.8, 6210.6, 6604.6 \\
Ti & 68 & 3998.6, 4534.8, 4548.8, 4798.5, 4865.6, 4874.0, 4911.2, 4913.6, 4915.2, 4926.1, 4964.7, 4981.4, 4981.7, 4997.1, 4999.5, 5005.1, 5009.6, 5016.2, 5024.8, 5039.9, 5043.6, 5113.4, 5129.1, 5145.5, 5147.5, 5152.2, 5185.9, 5210.4, 5211.5, 5219.7, 5224.3, 5295.8, 5336.8, 5366.6, 5381.0, 5384.6, 5418.8, 5426.2, 5460.5, 5490.1, 5490.7, 5490.8, 5662.1, 5689.5, 5702.6, 5716.4, 5866.4, 5903.3, 5918.5, 5922.1, 5937.8, 5953.2, 5965.8, 5978.5, 6064.6, 6091.2, 6092.8, 6126.2, 6258.1, 6261.1, 6312.2, 6336.1, 6554.2, 6556.1, 6599.1, 8683.0, 8692.3, 8734.7 \\
V & 35 & 3592.0, 3952.0, 4005.7, 4111.8, 4115.2, 4116.5, 4875.5, 5604.9, 5627.6, 5632.5, 5646.1, 5657.4, 5668.4, 5670.9, 5703.6, 5727.0, 5727.6, 5737.1, 6039.7, 6081.4, 6090.2, 6111.6, 6119.5, 6135.4, 6199.2, 6216.4, 6242.8, 6243.1, 6251.8, 6256.9, 6274.6, 6285.1, 6285.2, 6292.8, 6531.4 \\
Cr & 28 & 4588.2, 4646.2, 4789.3, 4848.2, 4936.3, 4953.7, 4964.9, 5200.2, 5206.0, 5239.0, 5241.4, 5246.8, 5247.6, 5272.0, 5287.2, 5296.7, 5300.7, 5304.2, 5312.9, 5318.8, 5345.8, 5348.3, 5409.8, 5628.6, 5719.8, 5788.4, 5838.7, 5844.6 \\
Mn & 13 & 4783.4, 4823.5, 5004.9, 5117.9, 5255.3, 5394.7, 5407.4, 5420.4, 5432.5, 5516.8, 6013.5, 6016.7, 6021.8 \\
Fe & 88 & 5044.2, 5054.6, 5127.4, 5127.7, 5197.6, 5198.7, 5225.5, 5234.6, 5242.5, 5247.1, 5250.2, 5264.8, 5295.3, 5322.0, 5325.6, 5373.7, 5379.6, 5386.3, 5414.1, 5466.4, 5467.0, 5522.4, 5546.5, 5560.2, 5577.0, 5618.6, 5636.7, 5638.3, 5650.0, 5651.5, 5661.3, 5679.0, 5701.5, 5705.5, 5855.1, 5905.7, 5927.8, 5934.7, 5956.7, 6003.0, 6005.5, 6024.1, 6027.1, 6056.0, 6065.5, 6079.0, 6082.7, 6084.1, 6093.6, 6096.7, 6151.6, 6157.7, 6165.4, 6173.3, 6188.0, 6200.3, 6213.4, 6219.3, 6226.7, 6240.6, 6247.6, 6252.6, 6265.1, 6270.2, 6271.3, 6380.7, 6432.7, 6456.4, 6498.9, 6593.9, 6597.6, 6625.0, 6678.0, 6703.6, 6705.1, 6710.3, 6713.7, 6725.4, 6726.7, 6733.2, 6750.2, 6752.7, 6806.8, 6810.3, 6837.0, 6839.8, 6843.7, 6858.1 \\
Co & 21 & 5176.1, 5212.7, 5280.6, 5301.0, 5331.4, 5352.0, 5369.6, 5483.3, 5483.9, 5530.8, 5590.7, 5647.2, 5915.6, 6093.1, 6117.0, 6189.0, 6429.9, 6455.0, 6632.4, 6771.0, 6814.9 \\
Ni & 26 & 3807.2, 4812.0, 4873.4, 4976.1, 4976.3, 4976.7, 5003.7, 5035.4, 5137.1, 5158.0, 5424.6, 5435.9, 5476.9, 5587.9, 5847.0, 6108.1, 6176.8, 6204.6, 6224.0, 6230.1, 6327.6, 6378.2, 6414.6, 6482.8, 6643.6, 6767.8 \\
Cu & 9 & 4022.6, 5105.5, 5153.2, 5218.2, 5220.1, 5700.2, 5782.1, 7933.1, 8092.6 \\
Zn & 4 & 4680.1, 4722.2, 4810.5, 6362.3 \\
Rb & 1 & 7947.6 \\
Sr & 11 & 4161.8, 4305.4, 4607.3, 4741.9, 4811.9, 4832.1, 4872.5, 4962.3, 6550.2, 6791.0, 7070.1 \\
Y & 27 & 3747.6, 3950.4, 3982.6, 4398.0, 4854.9, 4883.7, 4900.1, 4982.1, 5087.4, 5200.4, 5205.7, 5289.8, 5402.8, 5544.6, 5728.9, 6023.4, 6138.4, 6222.6, 6401.9, 6435.0, 6557.4, 6613.7, 6793.6, 6795.4, 6832.5, 6858.3, 7264.2 \\
Zr & 20 & 4040.2, 4050.3, 4090.5, 4161.2, 4258.1, 4317.3, 4443.0, 4497.0, 4784.9, 4809.4, 4828.0, 5112.3, 5350.1, 5350.4, 5385.1, 6127.4, 6127.5, 6134.6, 6140.5, 6143.2 \\
Nb & 5 & 3740.7, 4493.0, 4573.1, 5344.2, 7372.4 \\
Mo & 9 & 3864.1, 4279.1, 4474.6, 5506.5, 5533.0, 5570.4, 5751.4, 5858.3, 6030.7 \\
Rh & 1 & 3958.9 \\
Ba & 5 & 4130.6, 4166.0, 5853.7, 6141.7, 6496.9 \\
La & 32 & 3713.5, 4042.9, 4086.7, 4123.2, 4322.5, 4333.8, 4526.1, 4558.5, 4574.9, 4580.0, 4613.4, 4619.9, 4645.3, 4662.5, 4716.4, 4728.4, 4740.3, 4804.0, 4809.0, 4824.1, 4840.0, 4921.0, 4921.8, 4970.4, 4986.8, 5114.6, 5303.5, 5797.6, 5805.8, 5936.2, 6262.3, 6390.5 \\
Ce & 40 & 3960.9, 3992.4, 3999.2, 4003.8, 4042.6, 4053.5, 4073.5, 4075.7, 4083.2, 4115.4, 4118.1, 4120.8, 4137.6, 4142.4, 4145.0, 4165.6, 4289.9, 4382.2, 4391.7, 4418.8, 4449.3, 4460.2, 4471.2, 4483.9, 4523.1, 4539.7, 4562.4, 4572.3, 4593.9, 4628.2, 4773.9, 4914.9, 4984.4, 5044.0, 5117.2, 5187.4, 5274.2, 5330.6, 5353.5, 6043.4 \\
Pr & 10 & 3994.8, 4044.8, 4164.2, 4189.5, 4225.3, 4408.8, 5219.1, 5220.1, 5259.7, 5322.8 \\
Nd & 65 & 3780.4, 3784.2, 3784.8, 3990.1, 4004.0, 4013.2, 4018.8, 4021.3, 4023.0, 4043.6, 4051.2, 4060.0, 4069.3, 4109.4, 4133.4, 4156.1, 4211.3, 4214.6, 4232.2, 4303.6, 4446.4, 4463.0, 4542.6, 4567.6, 4579.3, 4594.4, 4645.8, 4706.5, 4715.6, 4811.3, 4825.5, 4849.1, 4859.0, 4914.4, 4947.0, 4959.1, 4970.9, 4989.9, 4998.5, 5089.8, 5092.8, 5130.6, 5143.3, 5200.1, 5212.4, 5234.2, 5249.6, 5250.8, 5255.5, 5276.9, 5293.2, 5306.5, 5311.5, 5319.8, 5357.0, 5361.5, 5416.4, 5533.8, 5548.4, 5581.6, 5619.0, 5740.9, 5842.4, 6365.5, 6637.2 \\
Sm & 22 & 3760.7, 3897.0, 3993.3, 4023.2, 4068.3, 4206.1, 4244.7, 4318.9, 4424.3, 4434.3, 4467.3, 4499.5, 4519.6, 4523.9, 4577.7, 4591.8, 4642.2, 4815.8, 4836.7, 4854.4, 4929.6, 4948.6 \\
Eu & 4 & 3819.7, 4129.7, 6049.5, 6645.1 \\
Gd & 8 & 3768.4, 3844.6, 4037.3, 4037.9, 4085.6, 4098.9, 4191.1, 4251.7 \\
Tb & 1 & 3702.8 \\
Dy & 6 & 3788.4, 3996.7, 4073.1, 4449.7, 4468.1, 5169.7 \\
Er & 7 & 3692.7, 3729.5, 3906.3, 3938.6, 4008.2, 4009.2, 4059.8 \\
Lu & 1 & 6221.9 \\
Hf & 1 & 4093.2 \\
W & 2 & 4843.8, 5224.7 \\
Os & 1 & 4135.8 \\
Pb & 3 & 4057.8 \\
\hline
\end{tabular}

\end{table*}

\begin{table*}
\centering
\footnotesize
\setlength{\tabcolsep}{4pt}
\caption{Preview of the Pb linelist, showing 20 lines nearest the strong Pb\,\textsc{i} components at 4057.801 and 4057.814\,\AA\. The full line list will be attached in machine-readable format.}
\label{tab:pb_lines}
\begin{tabular}{lccrl}
\hline
$\lambda$ (\AA) & Species & $\chi$ (eV) & $\log gf$ & Reference \\
\hline
\ldots & \ldots & \ldots & \ldots & \ldots \\
4057.794 & $^{51}$V\,\textsc{i} & 1.944 & $-4.126$ & \citet{Kurucz2011} \\
4057.794 & $^{51}$V\,\textsc{i} & 1.944 & $-3.871$ & \citet{Kurucz2011} \\
4057.797 & $^{51}$V\,\textsc{i} & 1.944 & $-5.779$ & \citet{Kurucz2011} \\
4057.797 & $^{51}$V\,\textsc{i} & 1.944 & $-3.871$ & \citet{Kurucz2011} \\
4057.797 & $^{51}$V\,\textsc{i} & 1.944 & $-3.672$ & \citet{Kurucz2011} \\
4057.801 & $^{12}$C$_2$ & 0.716 & $-5.951$ & \citet{Brooke2013} \\
4057.801 & $^{207}$Pb\,\textsc{i} & 1.320 & $-0.442$ & \citet{Biemont2000, Roederer2012} \\
4057.801 & $^{208}$Pb\,\textsc{i} & 1.320 & $-0.220$ & \citet{Biemont2000, Roederer2012} \\
4057.802 & $^{51}$V\,\textsc{i} & 1.944 & $-4.536$ & \citet{Kurucz2011} \\
4057.802 & $^{51}$V\,\textsc{i} & 1.944 & $-3.672$ & \citet{Kurucz2011} \\
4057.802 & $^{51}$V\,\textsc{i} & 1.944 & $-3.608$ & \citet{Kurucz2011} \\
4057.809 & $^{12}$CH & 1.760 & $-4.045$ & \citet{Masseron2014} \\
4057.809 & $^{12}$CH & 2.110 & $-7.777$ & \citet{Masseron2014} \\
4057.809 & $^{51}$V\,\textsc{i} & 1.944 & $-3.819$ & \citet{Kurucz2011} \\
4057.809 & $^{51}$V\,\textsc{i} & 1.944 & $-3.639$ & \citet{Kurucz2011} \\
4057.809 & $^{51}$V\,\textsc{i} & 1.944 & $-3.608$ & \citet{Kurucz2011} \\
4057.814 & $^{206}$Pb\,\textsc{i} & 1.320 & $-0.220$ & \citet{Biemont2000, Roederer2012} \\
4057.817 & $^{51}$V\,\textsc{i} & 1.944 & $-3.820$ & \citet{Kurucz2011} \\
4057.817 & $^{51}$V\,\textsc{i} & 1.944 & $-3.639$ & \citet{Kurucz2011} \\
4057.817 & $^{51}$V\,\textsc{i} & 1.944 & $-3.413$ & \citet{Kurucz2011} \\
\ldots & \ldots & \ldots & \ldots & \ldots \\
\hline
\end{tabular}
\end{table*}

Stellar parameter uncertainties come directly from \texttt{BACCHUS}. Final abundance uncertainties reflect the line-to-line scatter as well as the propagated parameter uncertainties.  To propagate parameter uncertainties, we perturbed the best-fit model atmosphere eight times, changing one parameter at a time by $\pm$ its uncertainty, and re-determined the abundances with each perturbed atmosphere. For each element, the shifts in the positive and negative directions were added separately in quadrature over the four parameters, giving asymmetric systematic uncertainties. These were then added in quadrature to the line-to-line standard deviation divided by $\sqrt{N-1}$ to get the final observed abundance uncertainty in element X: $\sigma_{\rm obs, X}$.  For single-line elements, we adopt a line-to-line standard deviation of 0.10. We require a minimum total uncertainty of 0.05 dex.

Our final adopted parameters and abundances are presented in Table \ref{tab:abundances}.

\begin{figure}
    \centering
    \includegraphics[width=\linewidth]{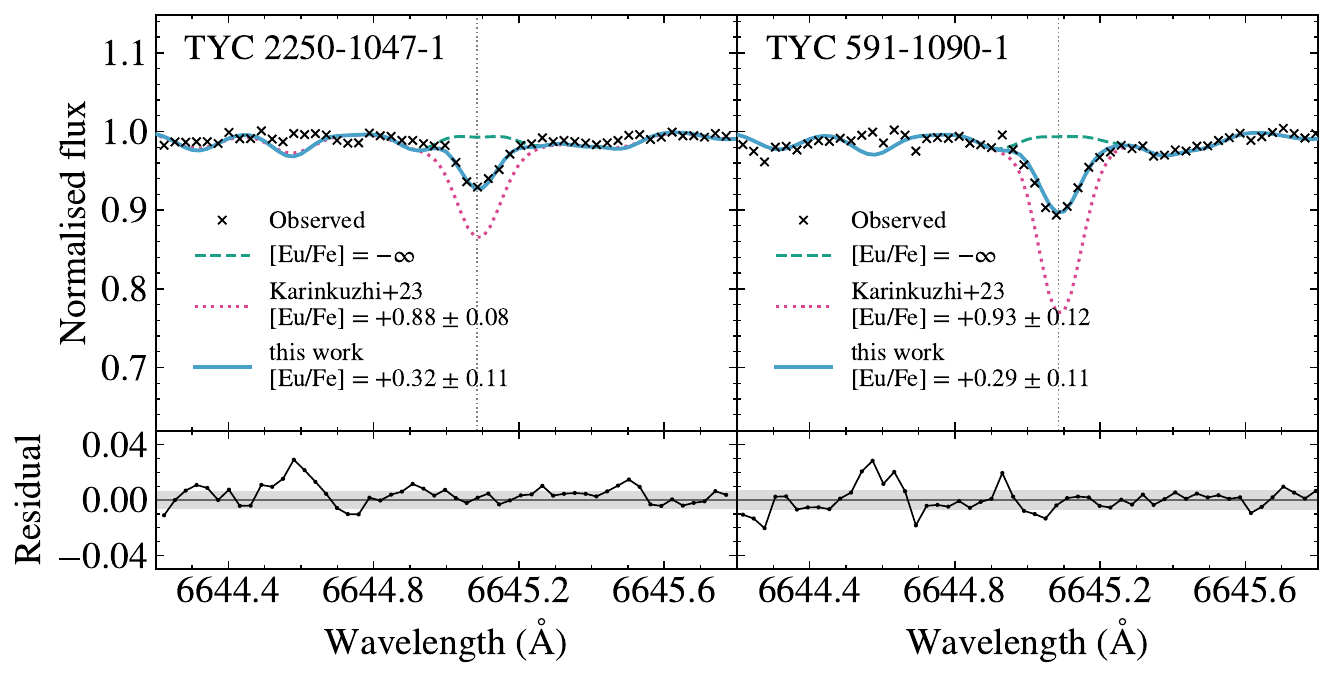}
    \caption{Best-fit synthetic spectra to the Eu \textsc{ii} $\lambda$6645.1 \AA~feature in archival UVES spectra of the two candidate \iprocess stars from K23.  We report much lower values of [Eu/Fe] that ultimately bring the abundance patterns of their stars into greater agreement with their theoretical \iprocess models. }
    \label{fig:eu}
\end{figure}


\begin{table*}
\centering
\caption{Parameters and abundances determined by this work. Numbers in parentheses report the number of lines used.}
\label{tab:abundances}
\begin{tabular}{lcccc}
\hline
 & UCAC4~278-068826 & UCAC4~179-210559 & UCAC4~265-014129 & UCAC4~445-009460  \\
\hline
Gaia DR3 & 6179152822938541696 & 6650507513536304384 & 5566309711910467328 & 3221200496767425024 \\
$T_\mathrm{eff}$ (K) & $4832\pm78$ & $5143\pm9$ & $6587\pm20$ & $5864\pm61$ \\
$\log g$ & $3.04\pm0.23$ & $3.76\pm0.22$ & $4.45\pm0.39$ & $4.63\pm0.29$ \\
$v_\mathrm{mic}$ (km\,s$^{-1}$) & $1.65\pm0.08$ & $1.03\pm0.03$ & $0.91\pm0.13$ & $1.25\pm0.11$ \\
{[Fe/H]} & $-0.70\pm0.05$ (70) & $-0.50\pm0.05$ (83) & $-0.32\pm0.05$ (73) & $-0.69\pm0.05$ (84) \\
\hline
A(Li) & \ldots & \ldots & $3.54\pm0.11$ (1) & \ldots \\
{[C/Fe]} & $+1.08\pm0.05$ (26) & $+0.48\pm0.05$ (29) & $+0.52\pm0.08$ (15) & $+1.01\pm0.05$ (34) \\
{[N/Fe]} & $+1.28\pm0.05$ (63) & $+0.27\pm0.05$ (51) & $+1.64\pm0.08$ (8) & $+0.76\pm0.05$ (55) \\
{[O/Fe]} & $+1.17\pm0.08$ (3) & $+0.65\pm0.06$ (3) & $-0.13\pm0.07$ (3) & $+0.65\pm0.05$ (3) \\
{[Na/Fe]} & $+0.25\pm0.10$ (5) & $+0.21\pm0.06$ (7) & $+0.18\pm0.08$ (8) & $+0.40\pm0.06$ (6) \\
{[Mg/Fe]} & $+0.42\pm0.07$ (7) & $+0.36\pm0.06$ (9) & $+0.36\pm0.07$ (11) & $+0.33\pm0.06$ (8) \\
{[Al/Fe]} & $+0.54\pm0.13$ (2) & $+0.37\pm0.05$ (2) & $+0.43\pm0.05$ (3) & $+0.36\pm0.11$ (1) \\
{[Si/Fe]} & $+0.28\pm0.06$ (12) & $+0.18\pm0.06$ (14) & $+0.41\pm0.06$ (14) & $+0.22\pm0.06$ (14) \\
{[S/Fe]} & $+0.61\pm0.11$ (1) & $+0.20\pm0.23$ (3) & \ldots & $+0.28\pm0.11$ (1) \\
{[K/Fe]} & \ldots & \ldots & $+1.16\pm0.11$ (1) & $+0.62\pm0.11$ (1) \\
{[Ca/Fe]} & $+0.15\pm0.07$ (13) & $+0.25\pm0.06$ (16) & $+0.61\pm0.06$ (24) & $+0.17\pm0.05$ (21) \\
{[Sc/Fe]} & $+0.38\pm0.07$ (9) & $+0.31\pm0.06$ (11) & $-0.02\pm0.13$ (5) & $+0.50\pm0.09$ (8) \\
{[Ti/Fe]} & $+0.25\pm0.05$ (49) & $+0.26\pm0.05$ (56) & $+0.40\pm0.06$ (33) & $+0.31\pm0.06$ (41) \\
{[V/Fe]} & $+0.06\pm0.05$ (27) & $+0.07\pm0.05$ (29) & $+0.58\pm0.08$ (9) & $+0.22\pm0.06$ (17) \\
{[Cr/Fe]} & $-0.13\pm0.09$ (8) & $-0.09\pm0.06$ (18) & $+0.65\pm0.07$ (16) & $+0.01\pm0.06$ (15) \\
{[Mn/Fe]} & $-0.68\pm0.14$ (3) & $-0.33\pm0.08$ (11) & $+0.62\pm0.07$ (10) & $-0.10\pm0.10$ (6) \\
{[Co/Fe]} & $+0.05\pm0.06$ (19) & $+0.07\pm0.06$ (21) & $+1.19\pm0.07$ (17) & $+0.23\pm0.08$ (7) \\
{[Ni/Fe]} & $+0.00\pm0.06$ (17) & $+0.01\pm0.06$ (22) & $+0.04\pm0.08$ (16) & $+0.14\pm0.06$ (21) \\
{[Cu/Fe]} & $+0.59\pm0.34$ (2) & $-0.08\pm0.07$ (6) & $+0.04\pm0.10$ (4) & $+0.25\pm0.07$ (6) \\
{[Zn/Fe]} & $+0.01\pm0.11$ (1) & $+0.09\pm0.06$ (3) & $-0.37\pm0.15$ (2) & $+0.31\pm0.07$ (4) \\
{[Rb/Fe]} & $+0.60\pm0.11$ (1) & $+0.26\pm0.11$ (1) & \ldots & \ldots \\
{[Sr/Fe]} & $+1.30\pm0.09$ (3) & \ldots & $+1.66\pm0.11$ (6) & $+1.68\pm0.09$ (6) \\
{[Y/Fe]} & $+1.22\pm0.08$ (15) & $+0.47\pm0.06$ (12) & $+0.44\pm0.08$ (14) & $+1.48\pm0.05$ (12) \\
{[Zr/Fe]} & $+1.31\pm0.09$ (11) & $+0.75\pm0.07$ (11) & $+0.61\pm0.12$ (6) & $+1.64\pm0.08$ (13) \\
{[Nb/Fe]} & $+1.18\pm0.27$ (2) & $+2.05\pm0.05$ (2) & $+2.31\pm0.11$ (1) & \ldots \\
{[Mo/Fe]} & $+1.14\pm0.07$ (4) & $+0.69\pm0.05$ (3) & \ldots & $+1.92\pm0.28$ (2) \\
{[Rh/Fe]} & \ldots & $+1.10\pm0.11$ (1) & \ldots & $+1.98\pm0.11$ (1) \\
{[Ba/Fe]} & $+1.69\pm0.11$ & $+0.94\pm0.16$ (2) & $+1.82\pm0.10$ (3) & $+1.89\pm0.11$ (1) \\
{[La/Fe]} & $+1.77\pm0.07$ (8) & $+1.08\pm0.07$ (18) & $+2.07\pm0.07$ (21) & $+2.00\pm0.06$ (24) \\
{[Ce/Fe]} & $+1.65\pm0.07$ (8) & $+0.94\pm0.06$ (26) & $+1.76\pm0.06$ (28) & $+1.87\pm0.06$ (21) \\
{[Pr/Fe]} & \ldots & $+0.80\pm0.06$ (5) & \ldots & $+1.85\pm0.06$ (5) \\
{[Nd/Fe]} & $+1.60\pm0.07$ (23) & $+0.94\pm0.05$ (50) & $+1.10\pm0.06$ (34) & $+1.87\pm0.05$ (53) \\
{[Sm/Fe]} & $+1.19\pm0.08$ (10) & $+0.72\pm0.06$ (15) & $+2.19\pm0.07$ (13) & $+1.63\pm0.06$ (17) \\
{[Eu/Fe]} & $+1.10\pm0.11$ (1) & $+0.59\pm0.11$ (1) & $+1.97\pm0.17$ (2) & $+1.32\pm0.11$ (1) \\
{[Gd/Fe]} & $+1.22\pm0.11$ (2) & $+0.75\pm0.08$ (2) & $+2.38\pm0.11$ (4) & $+1.35\pm0.06$ (5) \\
{[Tb/Fe]} & \ldots & \ldots & \ldots & $+1.37\pm0.11$ (1) \\
{[Dy/Fe]} & \ldots & $+1.57\pm0.11$ (1) & $+2.29\pm0.13$ (4) & \ldots \\
{[Er/Fe]} & $+1.76\pm0.30$ & $+0.63\pm0.11$ (1) & $+2.94\pm0.14$ (3) & $+1.75\pm0.05$ (3) \\
{[Lu/Fe]} & $+1.38\pm0.11$ (1) & \ldots & $+1.82\pm0.11$ (1) & \ldots \\
{[W/Fe]} & $+1.45\pm0.11$ (1) & $+1.02\pm0.08$ (2) & \ldots & \ldots \\
{[Pb/Fe]}$_\mathrm{LTE}$ & $+2.22\pm0.23$ (1)$^{a}$ & $+1.54\pm0.15$ (1)$^{a}$ & \ldots & $+1.88\pm0.25$ (1)$^{a}$ \\
{[Pb/Fe]}$_\mathrm{NLTE}$ & $+2.50\pm0.32$ (1)$^{a}$ & $+1.77\pm0.19$ (1)$^{a}$ & \ldots & $+2.14\pm0.27$ (1)$^{a}$ \\
\hline
{[Ba/Eu]} & $+0.59\pm0.16$ & $+0.35\pm0.19$ & $-0.15\pm0.20$ & $+0.57\pm0.16$ \\
{[La/Eu]} & $+0.67\pm0.13$ & $+0.49\pm0.13$ & $+0.10\pm0.18$ & $+0.68\pm0.13$ \\
{[Pb/Eu]}$_\mathrm{LTE}$ & $+1.12\pm0.25^{a}$ & $+0.95\pm0.19^{a}$ & \ldots & $+0.56\pm0.27^{a}$ \\
{[Pb/Eu]}$_\mathrm{NLTE}$ & $+1.40\pm0.34^{a}$ & $+1.18\pm0.22^{a}$ & \ldots & $+0.82\pm0.29^{a}$ \\
\hline
\end{tabular}
\begin{flushleft}
\footnotesize $^{a}$Measured by hand with spectral synthesis.
\end{flushleft}
\end{table*}

\begin{table*}
\centering
\addtocounter{table}{-1}
\caption{-- \textit{continued}}
\begin{tabular}{lcccc}
\hline
 & HD~216387 & TYC~1238-203-1  & TYC~2250-1047-1  & TYC~591-1090-1 \\
\hline
Gaia DR3 & 2603081134375972480 & 55149510384548992 & \ldots & \ldots \\
$T_\mathrm{eff}$ (K) & $6518\pm48$ & $5000\pm26$ & $5335\pm25^{b}$ & $5267\pm36^{b}$ \\
$\log g$ & $4.37\pm0.34$ & $3.53\pm0.05$ & $3.71\pm0.18^{b}$ & $3.68\pm0.50^{b}$ \\
$v_\mathrm{mic}$ (km\,s$^{-1}$) & $1.73\pm0.07$ & $1.21\pm0.05$ & $1.45\pm0.05^{b}$ & $1.18\pm0.06^{b}$ \\
{[Fe/H]} & $-0.28\pm0.05$ (71) & $-0.11\pm0.05$ (68) & $-0.59\pm0.05$ (75) & $-0.33\pm0.05$ (83) \\
\hline
{[C/Fe]} & $+0.50\pm0.06$ (42) & $+0.34\pm0.05$ (22) & $+0.27\pm0.05$ (26) & $+0.14\pm0.06$ (25) \\
{[N/Fe]} & $+1.26\pm0.11$ (1) & $+0.44\pm0.06$ (31) & \ldots & $+0.38\pm0.05$ (21) \\
{[O/Fe]} & $+0.64\pm0.08$ (3) & $+0.33\pm0.11$ (1) & $+0.39\pm0.05$ (2) & $+0.30\pm0.09$ (4) \\
{[Na/Fe]} & $+0.40\pm0.08$ (11) & $-0.23\pm0.09$ (3) & $+0.18\pm0.05$ (5) & \ldots \\
{[Mg/Fe]} & $+0.13\pm0.06$ (6) & $+0.11\pm0.08$ (3) & $+0.30\pm0.07$ (4) & $+0.18\pm0.08$ (7) \\
{[Al/Fe]} & $+0.17\pm0.16$ (2) & $+0.01\pm0.08$ (3) & $+0.25\pm0.05$ (2) & $+0.30\pm0.06$ (2) \\
{[Si/Fe]} & $+0.14\pm0.06$ (6) & $+0.01\pm0.06$ (9) & $+0.08\pm0.05$ (10) & $+0.05\pm0.06$ (11) \\
{[S/Fe]} & $+0.14\pm0.05$ (7) & $+0.04\pm0.11$ (1) & $-0.02\pm0.06$ (3) & $+0.28\pm0.11$ (1) \\
{[K/Fe]} & $+0.72\pm0.11$ (1) & \ldots & \ldots & $+0.28\pm0.11$ (1) \\
{[Ca/Fe]} & $+0.18\pm0.05$ (20) & $-0.01\pm0.06$ (11) & $+0.10\pm0.05$ (13) & $+0.15\pm0.07$ (17) \\
{[Sc/Fe]} & $+0.41\pm0.09$ (7) & $+0.02\pm0.08$ (11) & $+0.17\pm0.06$ (11) & $+0.10\pm0.07$ (10) \\
{[Ti/Fe]} & $+0.19\pm0.06$ (31) & $-0.10\pm0.05$ (46) & $+0.14\pm0.05$ (55) & $+0.06\pm0.06$ (51) \\
{[V/Fe]} & $+0.24\pm0.08$ (9) & $-0.19\pm0.05$ (23) & $+0.04\pm0.05$ (29) & $+0.01\pm0.06$ (23) \\
{[Cr/Fe]} & $-0.04\pm0.07$ (11) & $-0.24\pm0.05$ (16) & $-0.07\pm0.06$ (16) & $-0.19\pm0.07$ (18) \\
{[Mn/Fe]} & $-0.01\pm0.06$ (8) & $-0.19\pm0.05$ (5) & $-0.14\pm0.06$ (10) & $-0.49\pm0.11$ (7) \\
{[Co/Fe]} & $+0.30\pm0.07$ (5) & $-0.12\pm0.06$ (18) & $+0.08\pm0.06$ (20) & $-0.01\pm0.07$ (16) \\
{[Ni/Fe]} & $+0.05\pm0.06$ (17) & $-0.08\pm0.06$ (19) & $-0.01\pm0.05$ (22) & $+0.06\pm0.06$ (19) \\
{[Cu/Fe]} & $+0.18\pm0.08$ (6) & $-0.10\pm0.15$ (3) & $-0.02\pm0.10$ (4) & $+0.00\pm0.09$ (7) \\
{[Zn/Fe]} & $+0.09\pm0.09$ (3) & $-0.09\pm0.09$ (2) & $-0.01\pm0.15$ (2) & $-0.01\pm0.18$ (2) \\
{[Rb/Fe]} & \ldots & $+0.10\pm0.11$ (1) & \ldots & \ldots \\
{[Sr/Fe]} & $+1.21\pm0.18$ (3) & $+0.88\pm0.06$ (3) & $+1.02\pm0.15$ (3) & $+0.89\pm0.12$ (6) \\
{[Y/Fe]} & $+1.00\pm0.06$ (19) & $+0.72\pm0.07$ (13) & $+0.63\pm0.07$ (10) & $+0.74\pm0.07$ (18) \\
{[Zr/Fe]} & $+1.14\pm0.09$ (5) & $+0.77\pm0.06$ (11) & \ldots & \ldots \\
{[Nb/Fe]} & $+1.25\pm0.11$ (1) & \ldots & \ldots & \ldots \\
{[Mo/Fe]} & $+0.87\pm0.11$ (1) & $+0.84\pm0.07$ (4) & $+0.83\pm0.08$ (6) & $+1.87\pm0.11$ (1) \\
{[Ba/Fe]} & $+1.28\pm0.14$ (5) & \ldots & \ldots & $+0.88\pm0.35$ (2) \\
{[La/Fe]} & $+1.22\pm0.05$ (25) & $+0.83\pm0.08$ (16) & $+1.11\pm0.06$ (10) & $+0.96\pm0.07$ (20) \\
{[Ce/Fe]} & $+1.03\pm0.06$ (30) & $+0.66\pm0.06$ (14) & $+1.10\pm0.06$ (8) & $+0.85\pm0.07$ (14) \\
{[Pr/Fe]} & $+1.08\pm0.09$ (4) & $+0.54\pm0.06$ (4) & $+0.89\pm0.07$ (4) & $+0.66\pm0.08$ (7) \\
{[Nd/Fe]} & $+1.03\pm0.05$ (35) & $+0.54\pm0.05$ (30) & $+0.97\pm0.06$ (33) & $+0.90\pm0.06$ (34) \\
{[Sm/Fe]} & $+0.76\pm0.08$ (11) & $+0.35\pm0.07$ (6) & $+0.68\pm0.05$ (3) & $+0.75\pm0.08$ (11) \\
{[Eu/Fe]} & $+0.29\pm0.11$ (3) & $+0.32\pm0.11$ (1) & $+0.49\pm0.07$ (1) & $+0.44\pm0.09$ (1) \\
{[Gd/Fe]} & $+0.56\pm0.09$ (2) & \ldots & \ldots & \ldots \\
{[Tb/Fe]} & \ldots & $+0.00\pm0.11$ (1) & \ldots & \ldots \\
{[Dy/Fe]} & \ldots & \ldots & \ldots & $+0.89\pm0.14$ (2) \\
{[Er/Fe]} & $+0.85\pm0.11$ (1) & $+0.67\pm0.11$ (1) & \ldots & \ldots \\
{[Lu/Fe]} & $+0.80\pm0.11$ (1) & \ldots & \ldots & \ldots \\
{[Hf/Fe]} & \ldots & $+0.90\pm0.11$ (1) & \ldots & \ldots \\
{[Os/Fe]} & $+1.65\pm0.11$ (1) & \ldots & \ldots & \ldots \\
{[Pb/Fe]}$_\mathrm{LTE}$ & $<+1.27^{a}$ & $<-0.08^{a}$ & \ldots & \ldots \\
{[Pb/Fe]}$_\mathrm{NLTE}$ & $<+1.46^{a}$ & $<+0.09^{a}$ & \ldots & \ldots \\
\hline
{[Ba/Eu]} & $+0.99\pm0.18$ & \ldots & \ldots & $+0.44\pm0.36$ \\
{[La/Eu]} & $+0.93\pm0.12$ & $+0.51\pm0.14$ & $+0.62\pm0.09$ & $+0.52\pm0.11$ \\
{[Pb/Eu]}$_\mathrm{LTE}$ & $<+0.98^{a}$ & $<-0.40^{a}$ & \ldots & \ldots \\
{[Pb/Eu]}$_\mathrm{NLTE}$ & $<+1.17^{a}$ & $<-0.23^{a}$ & \ldots & \ldots \\
\hline
\end{tabular}
\begin{flushleft}
\footnotesize $^{a}$Measured by hand with spectral synthesis.\\
$^{b}$Adopted from \citet{Karinkuzhi2021b, Karinkuzhi2023}.
\end{flushleft}
\end{table*}

\section{Isotopic ratios}
\label{sec:isotope_methods}



We synthesised $^{12}$C/$^{13}$C via the CN red system around 8000\,\AA, a region previously used in the literature for carbon isotope measurements \cite[e.g.,][]{Briley1997AJ....113..306B, BocekTopcu2015MNRAS.446.3562B,Takeda2019PASJ...71..119T, Shejeelammal2021MNRAS.502.1008S, Aguilera2023A&A...670A..73A}. The CN line list was constructed by A.~McWilliam (priv.\ comm.), after comparison with the ExoMol KTPSYT-iso line list \citep{Kozlov2025}. We further supplemented this list with \texttt{linemake} \citep{linemake}, which draws on the CN line list of \citet{Sneden2014ApJS..214...26S}:
laboratory wavelengths where available, and transition probabilities from the calculations of \citet{Brooke2014ApJS..210...23B}. Model atmospheres were MARCS models \citep{Gustafsson2008} interpolated with \texttt{q2} \citep{Ramirez2014} to be read in by MOOG \citep{Sneden1973}. We used the 2017 version of MOOG that includes scattering \citep{Sobeck2011AJ....141..175S}, as distributed by A.~Ji (\texttt{moog17scat}\footnote{\url{https://github.com/alexji/moog17scat}}).

Syntheses were computed on a grid of $\varepsilon$(N), $^{13}$C fraction
($f_{13} = {}^{13}\mathrm{C}/({}^{12}\mathrm{C}+{}^{13}\mathrm{C})$) and macroturbulent
velocity, convolved with a Gaussian matching the instrumental resolving power, and interpolated during the fit. Rather than restricting the fit to the commonly used 8000--8010\,\AA\ synthesis region, we used a wavelength range of 8000.5--8016\,\AA\ to increase the number of available $^{13}$CN features. This reduced the sensitivity of the derived ratios to blending, telluric contamination or line-list errors affecting any individual feature, and also improved continuum placement. Using the \texttt{dynesty} nested sampler \citep{Speagle2020MNRAS.493.3132S} we fit the free parameters $f_{13}$, $\varepsilon$(N), $v_\mathrm{mac}$, the radial velocity and the pixel noise. We modelled the continuum as a piecewise-linear function with knots every 4\,\AA\ and marginalised it analytically, so uncertainty in its placement propagates into the derived ratio. To account for correlated residuals between adjacent pixels, the log-likelihood was divided by the effective number of pixels per independent residual, estimated from the best-fit residuals by blocking. We adopt a uniform prior on the $^{13}$C fraction, $f_{13} = 1/(1+^{12}\mathrm{C}/^{13}\mathrm{C})$, over $10^{-4} < f_{13} < 0.35$, corresponding to $^{12}$C/$^{13}$C $\approx 2$–$10^4$. We report the 16th and 84th posterior percentiles as uncertainties, and the 5th percentile as a 95\% lower limit where the upper bound is poorly constrained. To assess sensitivity to the adopted prior, we repeat the fits using a prior uniform in $\log(^{12}\mathrm{C}/^{13}\mathrm{C})$. For UCAC4 278-068826, both priors yield consistent, well-bounded posterior distributions, with median ratios of 67 and 69, respectively. We therefore report a measurement for this star, while adopting conservative lower limits for the remaining two stars with $^{13}\mathrm{C}$ features, whose upper bounds are strongly prior-dependent. The quoted uncertainties are statistical and do not include errors in the adopted stellar parameters or C and O abundances.

We used Arcturus spectra (from the \citealt{Hinkle2000} atlas) to verify the line list and fitting method, adopting the stellar parameters of \citet{Ramirez2011ApJ...743..135R} and the C and O
abundances of \citet{Sneden2014ApJS..214...26S}. We derive an isotopic ratio of $^{12}$C/$^{13}$C $= 7.7 \pm 0.5$, in agreement with literature values \citep{Sneden2014ApJS..214...26S}. We also tested the line list on the Sun, using the solar flux spectrum from the same atlas. However, as the Sun
has little $^{13}$C, this test validated only the locations and strengths of the $^{12}$CN and atomic lines.

We note an inconsistency in the line list around 8011.5\,\AA, seen in both Arcturus and the Sun: the observed absorption lies $\sim$0.3\,\AA\ blueward of the predicted position of a $^{12}$CN line, adjacent to the $^{13}$CN features at 8011.2\,\AA. 
We attribute the discrepancy to an inaccurate predicted wavelength for this line, likely computed from molecular constants beyond the range of laboratory measurements. We masked this region (8011.22--8011.70\,\AA) and did not use it to determine isotopic ratios. Two further regions poorly reproduced in both Arcturus and the Sun (8006.25--8006.85 and 8014.10--8014.30\,\AA) were masked for all stars.

We identified tellurics by cross-correlating each spectrum with the telluric spectrum from the \citet{Hinkle2000} Arcturus atlas to determine the telluric template shift. Points where the template absorption is deeper than 1\% of the continuum (padded by $\pm0.06$\,\AA) were removed from the likelihood.

For the program stars we adopt the stellar parameters derived in this study, and replace the scaled-solar C and O abundances with the values listed in Table \ref{tab:abundances}. We treat N as a free parameter to reproduce the CN absorption features. We also iteratively remove points more than $3\sigma$ from the best fit and refit, to minimise the influence of inaccuracies in the line list, enhancements of $s-$process elements above scaled-solar values, and residual telluric contamination. Pixels within $\pm0.08$\,\AA\ of the $^{13}$CN features and of the twelve strongest $^{12}$CN lines in each star were never removed, so that the isotopic information is retained. 

For Mg isotopes, we use the same methodology developed in \cite{McKenzie2024MNRAS.527.7940M, McKenzie2024MNRAS.534L..35M}, which uses an architecture similar to that discussed for C isotopes. Minor alterations to the code were made to account for the significant C$_2$ and \sprocess enhancement within these stars. For UCAC4 179-210559 we assume a $^{12}$C/$^{13}$ of 50, which is consistent with its upper-limit measurement and stellar evolution. For TYC 2250-1047-1, for which there is no $^{12}$C/$^{13}$ measurement, we assume a $^{12}$C/$^{13}$ of 20; however, we do not expect the $^{12}$C/$^{13}$ ratio to make any statistically significant difference to our Mg isotopic ratio. We explicitly set the abundances of C, Cr, Fe, Y, Zr, Nb, La, Ce, Pr, Nd, Sm, Eu, Dy, and W to account for blended features. 

We use the same regions to derive Mg isotopes from the molecular MgH lines (listed in Table \ref{tab:mg_isotopes}), and we refer readers to the original studies for further discussion of the specific methodology and line list windows.

\begin{table*}
\centering
\caption{$^{12}$C/$^{13}$C ratios from CN red-system synthesis over 8000.5--8016\,\AA.}
\label{tab:c12c13}
\begin{tabular}{llp{10cm}}
\hline
Star & $^{12}$C/$^{13}$C & Notes \\
\hline
UCAC4 278-068826 & $66^{+16}_{-11}$ & $^{13}$CN detected; the ratio is bounded and insensitive to the choice of prior. \\
TYC 1238-203-1          & $>19$            & Lower limit; negligible $^{13}$C contribution cannot be excluded. \\
UCAC4 179-210559 & $>18$            & Lower limit; posterior is consistent with $f_{13} = 0$. \\
UCAC4 265-014129        & \ldots           & CN not detected. \\
UCAC4 445-009460 & \ldots           & Unconstrained; $^{13}$CN features blended with telluric absorption. \\
HD 216387            & \ldots           & CN not detected. \\
TYC 2250-1047-1          & \ldots           & No spectral coverage at 8000\,\AA. \\
TYC 591-1090-1   & \ldots           & Unconstrained; S/N too low. \\
\hline
\end{tabular}
\begin{flushleft}
\footnotesize
\textit{Notes.} The measurement gives the posterior median and 68 per cent interval; limits are one-sided 95 per cent lower limits. All values assume a prior uniform in $f_{13} = {}^{13}\mathrm{C}/({}^{12}\mathrm{C}+{}^{13}\mathrm{C})$. The continuum placement, $\varepsilon$(N), $v_\mathrm{mac}$ and radial velocity were fitted simultaneously with $f_{13}$, and their uncertainties are propagated into the quoted values.
\end{flushleft}
\end{table*}

\begin{table*}
\centering
\caption{Magnesium isotopic abundances derived from MgH spectral synthesis following the methodology in \protect\cite{McKenzie2024MNRAS.527.7940M, McKenzie2024MNRAS.534L..35M}.}
\label{tab:mg_isotopes}
\begin{tabular}{lcccccc}
\hline
Star &
$^{25}$Mg/$^{24}$Mg &
$^{26}$Mg/$^{24}$Mg &
$^{24}$Mg (\%) &
$^{25}$Mg (\%) &
$^{26}$Mg (\%) &
Regions used \\
\hline

UCAC4 179-210559 &
$0.031 \pm 0.022$ &
$0.032 \pm 0.020$ &
$94.13 \pm 3.58$ &
$2.90 \pm 1.86$ &
$2.97 \pm 1.72$ &
1, 2, 3, 7, 8, 9, 10 \\

TYC 2250-1047-1 &
$0.097 \pm 0.061$ &
$0.110 \pm 0.058$ &
$82.82 \pm 7.41$ &
$8.07 \pm 3.85$ &
$9.11 \pm 3.56$ &
2, 3, 4, 7, 8, 9, 10 \\

UCAC4 278-068826 &
\multicolumn{3}{l}{Too much C$_2$ contamination} &
\ldots \\

UCAC4 445-009460 &
\multicolumn{3}{l}{Too much C$_2$ contamination} &
\ldots \\

UCAC4 265-014129 &
\multicolumn{3}{l}{MgH features too weak to measure} &
\ldots \\

HD 216387 &
\multicolumn{3}{l}{MgH features too weak to measure} &
\ldots \\

TYC 591-1090-1 &
\multicolumn{3}{l}{Insufficient S/N} &
\ldots \\

TYC 1238-203-1 &
\multicolumn{3}{l}{Too much C$_2$ contamination} &
\ldots \\

\hline
\end{tabular}
\end{table*}

\section{Comparison to Models and Final Best-Fits}\label{appendix:models}

For each star, we collect all models within 0.20 dex of the star's [Fe/H].  These models represent the final AGB surface composition at the end of the thermally pulsing phase.  However, only a fraction of this material is transferred to the present-day barium star and mixed into its envelope.  As such, we must dilute the AGB abundance pattern to match the observed enrichment.  We parameterise this dilution with $\delta$, the mass fraction of the barium star’s mixed envelope that is accreted AGB material (as in \citealt[e.g.,][]{Cseh2022}), following their methodology with two modifications.  Firstly, we anchor our $\delta$ to [La/Fe], our most reliably measured element which exhibits strong yet unsaturated lines.  Secondly, rather than adopting a solar-scaled initial composition, we assume an initial composition that scales with Galactic disk trends.  While predominantly \sprocess elements show little [X/Fe] evolution in the metallicity range of our sample, \rprocess elements like Eu increase toward lower metallicity \citep[][Manea et al.,submitted]{Mishenina2022}.  To account for this evolution, we split each element into its $s$- and $r$-process fractions in the solar system, $f_s$ and $1-f_s$ \citep{Prantzos2020}. We keep the $s$-process part solar-scaled and enhance the $r$-process part following a linear disk [Eu/Fe] trend, (anchored to $+0.28$ at [Fe/H]~$=-0.7$ and solar at solar metallicity).  This avoids crediting typical disk evolution of Eu to the companion, which could bias the more metal-poor stars towards the \iprocess model.  We test our analysis on the pure solar initial composition assumption, however, and our results do not qualitatively change.

With these modifications, the final adjusted [X/Fe] model values to which we directly compare our observed abundances are:
\begin{equation}
[{\rm X/Fe}]_{\rm model, adjusted} = \log_{10}\left((1-\delta)\,10^{[{\rm X/Fe}]_{0}} + \delta\,10^{[{\rm X/Fe}]_{\rm model, original}}\right)
\label{eq:dilution}
\end{equation}
where the natal stellar abundance $[{\rm X/Fe}]_{0}$ is estimated with:
\begin{equation}
[{\rm X/Fe}]_{0} = \log_{10}\left(f_s + (1-f_s)\,10^{[{\rm Eu/Fe}]_{\rm disk}}\right)
\label{eq:x0}
\end{equation}
and $\delta$ is set by requiring the model to reproduce the observed [La/Fe]:
\begin{equation}
\delta = \frac{10^{[{\rm La/Fe}]_{\rm obs}} - 10^{[{\rm La/Fe}]_{0}}}{10^{[{\rm La/Fe}]_{\rm AGB}} - 10^{[{\rm La/Fe}]_{0}}}
\label{eq:d}
\end{equation}
As in \citet{Cseh2022}, we reject all models where $\delta$ exceeds 0.90, which would imply that the barium star’s envelope is almost entirely accreted AGB material, requiring more mass than the companion could plausibly have transferred.

Finally, we compute $\chi^2$ between our abundances and each viable model:
\begin{equation}
\chi^2 = \sum_{X} \frac{\left([{\rm X/Fe}]_{\rm obs} - [{\rm X/Fe}]_{\rm model, adjusted}\right)^2}{\sigma_{{\rm obs},X}^2 + \sigma_{\rm model}^2 + \left(f_{0,X}\,\sigma_{0,X}\right)^2},
\end{equation}
where $\sigma_{\rm model}$ is the model uncertainty assumed to be 0.05 dex, $\sigma_{0,X}$ is the uncertainty in the initial stellar composition, which we set to 0.1 to reflect the intrinsic abundance dispersion of disk stars (e.g., Manea et al., submitted), and $f_{0,X}$ is the fraction of the model's final adjusted composition that comes from the initial stellar composition:

\begin{equation}
f_{0,X} = \frac{(1-\delta)\,10^{[{\rm X/Fe}]_{0}}}{10^{[{\rm X/Fe}]_{\rm model, adjusted}}},
\end{equation}

We only consider the elements from La to Pb in our fit.  We exclude the lighter elements (Rb-Rh) because \iprocess models cannot reproduce them reliably.  K23 found that Sr, Y and Zr are poorly matched by both their \sprocess and \iprocess models and dominate their $\chi^2$. Furthermore, we exclude Ba from our fits.  At these extreme abundances and high metallicities, Ba is severely saturated and difficult to model spectroscopically, something that is reflected in our large uncertainties.  Furthermore, we exclude Er because it is measured from the $\lambda$4009\AA~line that is blended with Ce at these extreme enrichment levels and can skew fits towards higher Er, creating a preference for \iprocess.  Dy and Os are also excluded because they return anomalously high abundances that are not predicted by any model for most stars and do not meaningfully distinguish between the two models.  However, we note that including these elements does not change our conclusions qualitatively.  For each $\chi^2$, we determine the associated $p$-values using \texttt{scipy.stats.chi2.sf}, with $\nu = N_{\rm fit} - 1$ degrees of freedom, and we consider a model able to explain a star's abundance pattern if $p > 0.01$.  We note once again that these $\chi^2$ and $p$-values assume uncorrelated Gaussian uncertainties and should be considered approximate \citep[see discussion in][]{Hampel2019}.

Our best-fitting models and their associated parameters are detailed in Table \ref{tab:models}.

\begin{table*}
\centering
\setlength{\tabcolsep}{4pt}
\caption{Best-fitting AGB models for each star.}
\label{tab:models}
\begin{tabular}{lc|cc|cc|cccccc}
\hline
 & & \multicolumn{2}{c|}{K23 $i$-process} & \multicolumn{2}{c|}{K23 $s$-process} & \multicolumn{6}{c}{Best $s$-process model (Monash/FRUITY)} \\
Star & $N_{\rm fit}$ & $\chi^2_\nu$ & $\delta$ & $\chi^2_\nu$ & $\delta$ & Grid & $M$ & $Z$ & Details & $\chi^2_\nu$ & $\delta$ \\
 & & & & & & & ($M_\odot$) & & & & \\
\hline
UCAC4 278-068826 & 9 & 1.39 & 0.025 & 16.26 & 0.240 & Monash & 2 & 0.0028 & $M_{\rm mix}=6\times10^{-3}\,M_\odot$ & 3.20 & 0.248 \\
UCAC4 179-210559 & 9 & 1.18 & 0.005 & 8.76 & 0.045 & FRUITY & 1.5 & 0.003 & $v_{\rm rot}=0$\,km\,s$^{-1}$, extended & 2.90 & 0.206 \\
UCAC4 265-014129$^{a}$ & 7 & 39.70 & 0.050 & 79.75 & 0.479 & K26-s & 2 & \ldots & \ldots & 79.75 & 0.479 \\
UCAC4 445-009460 & 9 & 2.18 & 0.043 & 16.50 & 0.411 & Monash & 2 & 0.0028 & $M_{\rm mix}=6\times10^{-3}\,M_\odot$ & 1.66 & 0.424 \\
HD 216387 & 9 & 8.08 & 0.007 & 4.55 & 0.065 & FRUITY & 2 & 0.01 & $v_{\rm rot}=0$\,km\,s$^{-1}$, extended & 1.04 & 0.422 \\
TYC 1238-203-1 & 9 & 14.36 & 0.002 & 4.61 & 0.024 & Monash & 1.5 & 0.014 & $M_{\rm mix}=2\times10^{-3}\,M_\odot$ & 2.38 & 0.705 \\
TYC 2250-1047-1 & 6 & 0.43 & 0.005 & 2.18 & 0.048 & FRUITY & 1.5 & 0.003 & $v_{\rm rot}=60$\,km\,s$^{-1}$, extended & 1.52 & 0.734 \\
TYC 591-1090-1 & 7 & 1.14 & 0.004 & 4.70 & 0.033 & FRUITY & 2 & 0.008 & $v_{\rm rot}=0$\,km\,s$^{-1}$, standard & 1.63 & 0.264 \\
\hline
\end{tabular}
\begin{flushleft}
\footnotesize
\textit{Notes.} $\chi^2_\nu$ is the reduced $\chi^2$ over the elements La--Pb used in the fit. $\delta$ is the mass fraction of AGB material in the mixed envelope ($\delta = 1/\mathrm{dil}$ in \citealt{Cseh2022}), calibrated to the observed [La/Fe]. The K23 models are for a $2\,M_\odot$, [Fe/H] $=-0.5$ AGB star \citep{Karinkuzhi2023}. Additional $s$-process models are considered from Monash \citep{Karakas2016, Karakas2018} and FRUITY \citep{Cristallo2011, Cristallo2015} grids within $\pm0.2$\,dex of the star's [Fe/H].\\
$^{a}$Abundance pattern entirely inconsistent with any AGB model.
\end{flushleft}
\end{table*}

\bsp	
\label{lastpage}
\end{document}